\documentclass[a4paper,aps,prc,twocolumn,superscriptaddress,nobibnotes,showpacs]{revtex4-1}
\usepackage[utf8x]{inputenc}
\usepackage[english]{babel}
\usepackage{fontenc}
\usepackage{graphicx}
\usepackage{color}
\usepackage{textcomp}
\usepackage{multirow}
\usepackage{todonotes}
\usepackage{hyperref}
\usepackage{amsmath,amssymb,amsbsy}

\begin{document}

\title{Examining the $N$ = 28 shell closure \\ through high-precision mass measurements of $^{46−48}$Ar}


\author{M. Mougeot}
\email[Corresponding author: ]{maxime.mougeot@cern.ch}
\altaffiliation{Present Address: CERN, 1211 Geneva, Switzerland}
\affiliation{CSNSM-IN2P3-CNRS, Universit\'{e} Paris-Sud, 91405 Orsay, France}

\author{D. Atanasov}
\altaffiliation{Present Address: KU Leuven, Instituut voor Kern- en Stralingsfysica, 3001 Leuven, Belgium}
\affiliation{Max-Planck-Institut f\"{u}r Kernphysik, Saupfercheckweg 1, 69117 Heidelberg, Germany}
\affiliation{Technische Universit\"{a}t Dresden, 01069 Dresden, Germany}

\author{C. Barbieri}
\affiliation{Department of Physics, University of Surrey, Guildford GU2 7XH, UK}

\author{K. Blaum}
\affiliation{Max-Planck-Institut f\"{u}r Kernphysik, Saupfercheckweg 1, 69117 Heidelberg, Germany}

\author{M. Breitenfeld}
\affiliation{CERN, 1211 Geneva, Switzerland}

\author{A. de Roubin}
\altaffiliation{Present Address: The University of Jyv\"{a}skyl\"{a}, Department of Physics, PO Box 35 (YFL), FIN-40014, Finland}
\affiliation{Max-Planck-Institut f\"{u}r Kernphysik, Saupfercheckweg 1, 69117 Heidelberg, Germany}

\author{T. Duguet}
\affiliation{IRFU, CEA, Universit\'e Paris-Saclay, 91191 Gif-sur-Yvette, France} 
\affiliation{KU Leuven, Instituut voor Kern- en Stralingsfysica, 3001 Leuven, Belgium}

\author{S. George}
\affiliation{Max-Planck-Institut f\"{u}r Kernphysik, Saupfercheckweg 1, 69117 Heidelberg, Germany}

\author{F. Herfurth}
\affiliation{GSI Helmholtzzentrum f\"{u}r Schwerionenforschung GmbH, Planckstra\ss e 1, 64291 Darmstadt, Germany}

\author{A. Herlert}
\affiliation{FAIR GmbH, Planckstra\ss e 1, 64291 Darmstadt, Germany}

\author{J.D. Holt}
\affiliation{TRIUMF, 4004 Wesbrook Mall, Vancouver BC V6T 2A3, Canada}

\author{J. Karthein}
\affiliation{Max-Planck-Institut f\"{u}r Kernphysik, Saupfercheckweg 1, 69117 Heidelberg, Germany}
\affiliation{CERN, 1211 Geneva, Switzerland}

\author{D.~Lunney}
\affiliation{CSNSM-IN2P3-CNRS, Universit\'{e} Paris-Sud, 91405 Orsay, France}

\author{V. Manea}
\affiliation{Max-Planck-Institut f\"{u}r Kernphysik, Saupfercheckweg 1, 69117 Heidelberg, Germany}
\affiliation{CERN, 1211 Geneva, Switzerland}

\author{P. Navr\'atil}
\affiliation{TRIUMF, 4004 Wesbrook Mall, Vancouver BC V6T 2A3, Canada}

\author{D. Neidherr}
\affiliation{GSI Helmholtzzentrum f\"{u}r Schwerionenforschung GmbH, Planckstra\ss e 1, 64291 Darmstadt, Germany}

\author{M. Rosenbusch}
\altaffiliation{Present Address: RIKEN Nishina Center for Accelerator-Based Science, Wako, Saitama 351-0198, Japan
}
\affiliation{Universit\"{a}t Greifswald, Institut f\"{u}r Physik, 17487 Greifswald, Germany}

\author{L. Schweikhard}
\affiliation{Universit\"{a}t Greifswald, Institut f\"{u}r Physik, 17487 Greifswald, Germany}

\author{A.~Schwenk}
\affiliation{Institut f\"ur Kernphysik, Technische Universit\"at Darmstadt, 64289 Darmstadt, Germany}
\affiliation{ExtreMe Matter Institute EMMI, GSI Helmholtzzentrum f\"ur Schwerionenforschung GmbH, 64291 Darmstadt, Germany}
\affiliation{Max-Planck-Institut f\"{u}r Kernphysik, Saupfercheckweg 1, 69117 Heidelberg, Germany}

\author{V. Som\`{a}}
\affiliation{IRFU, CEA, Universit\'e Paris-Saclay, 91191 Gif-sur-Yvette, France} 

\author{A. Welker}
\affiliation{Technische Universit\"{a}t Dresden, 01069 Dresden, Germany}
\affiliation{CERN, 1211 Geneva, Switzerland}

\author{F. Wienholtz}
\altaffiliation{Present Address: Institut f\"ur Kernphysik, Technische Universit\"at Darmstadt, 64289 Darmstadt, Germany}
\affiliation{Universit\"{a}t Greifswald, Institut f\"{u}r Physik, 17487 Greifswald, Germany}
\affiliation{CERN, 1211 Geneva, Switzerland}

\author{R.N. Wolf}
\affiliation{Max-Planck-Institut f\"{u}r Kernphysik, Saupfercheckweg 1, 69117 Heidelberg, Germany}
\affiliation{ARC Centre of Excellence for Engineered Quantum Systems,The University of Sydney, NSW 2006, Australia}

\author{K. Zuber}
\affiliation{Technische Universit\"{a}t Dresden, 01069 Dresden, Germany}

\date{\today}

\begin{abstract}
The strength of the $N$ = 28 magic number in neutron-rich argon isotopes is examined through high-precision mass measurements of $^{46-48}$Ar, performed with the ISOLTRAP mass spectrometer at ISOLDE/CERN. The new mass values are up to 90 times more precise than previous measurements. While they suggest the persistence of the $N$ = 28 shell closure for argon, we show that this conclusion has to be nuanced in light of the wealth of spectroscopic data and theoretical investigations performed with the \emph{SDPF-U} phenomenological shell model interaction. Our results are also compared with \emph{ab initio} calculations using the Valence Space In-Medium Similarity Renormalization Group and the Self-Consistent Green's Function approaches.
Both calculations provide a very good account of mass systematics at and around $Z$ = 18 and, generally, a consistent description of the physics in this region.
This combined analysis indicates that $^{46}$Ar is the transition between the closed-shell $^{48}$Ca and collective $^{44}$S.
\end{abstract}


\maketitle

\section{Introduction}

Just as the experimental evidence for ``magic" proton and neutron numbers was instrumental for laying a basic foundation of nuclear theory \cite{PhysRev.75.1969,PhysRev.75.1766.2}, the observation of their demise in exotic nuclear systems \cite{PhysRevC.12.644} was pivotal for the establishment of the modern understanding of nuclear structure and the mechanisms driving its evolution far from $\beta$-stability \cite{PhysRevLett.87.082502,Sorlin2008602,Smirnova2010109}. 
The magic numbers found their origin in systematic studies of mass differences \cite{elsasser}. The disappearance of the magic $N=20$ shell closure was likewise evidenced through mass measurements of exotic sodium ($Z=11$) isotopes \cite{PhysRevC.12.644}, for which the binding energy normally reduced beyond a shell closure was in fact found to increase due to deformation. This was attributed to intruder configurations forming what is now known as the ``island of inversion” \cite{ioi}.

The question of the persistence of the next magic number -- $N = 28$ -- below the doubly magic (stable) $^{48}$Ca isotope has been subjected to detailed experimental scrutiny over the past two decades. The demise of the $N$ = 28 spherical gap in the silicon ($Z$ = 14) isotopic chain has been established through various spectroscopic studies \cite{Grevy2003,Bastin2007,Campbell2006,Takeuchi2012,Stroberg2014,PhysRevLett.122.222501} while the sulfur chain ($Z$ = 16) 
shows signatures of shape-coexistence in the vicinity of $^{44}$S \cite{Glasmacher1997,Gade2009,Gaudefroy2009,Force2010,Santiago-Gonzalez2011,PhysRevC.100.044312}, a phenomenon often encountered at the border of an island of inversion \cite{PhysRevLett.117.272501}. 

The argon ($Z$ = 18) chain is however less clear cut. 
A relatively healthy $N$ = 28 gap is attested by the high lying E(2$_{1}^{+}$) excitation energy \cite{Gade2009,Bhattacharyya2008}, which is one of the major indicators of a closed shell. The level scheme proposed for $^{45}$Ar in \cite{Grevy2003} was also found to be well described in a single-particle picture and little collectivity. Likewise, investigations of neutron-rich argon isotopes via neutron knockout reactions \cite{Gade2005} portray $^{46}$Ar as a seemingly ``good" semi-magic nuclide with a low observed cross section to the $3 / 2^{-}$ state in $^{45}$Ar. However, later results from $(d,p)$ transfer reactions performed at GANIL \cite{Gaudefroy2006,Gaudefroy2008} hinted at the erosion of the $N$ = 28 shell gap already at $Z$ = 18. Good indicators of the onset of collective nuclear behavior, B(E2: 2$_{1}^{+} \rightarrow$0$_{1}^{+}$) values also give conflicting results. Three independent measurements yield a rather low B(E2) value \cite{Scheit1996,Gade2003,Calinescu2014,Winkler2012} compatible with the persistence of the $N$ = 28 gap in this chain, while the B(E2) extracted from a life-time measurement \cite{Mengoni2010}, albeit a low statistics one, suggests the opposite.

Ground-state properties provide complementary and model-independent probes of nuclear phenomena. Laser-spectroscopy measurements of mean-square charge radii along the argon isotopic chain show a pronounced shell effect at $N$ = 28 \cite{KLEIN19961,BLAUM200830}. Likewise, mass measurements performed using the S800 spectrometer at the NSCL suggest the presence of a strong $N$ = 28 shell in the argon chain \cite{Meisel2015}, but the large uncertainties of these masses prevent from making definitive statements.  
Mass measurements of the $N=28$ gap below calcium 
\cite{PhysRevLett.84.5062,Jurado2007,Ringle2009} hint at its possible erosion 
for chlorine ($Z=17$) and sulfur ($Z=16$) but again, no firm conclusions can be drawn due to the experimental uncertainties beyond $N=28$.    

Neutron-rich nuclei in this region of deformation below $Z$ = 20 are also of great theoretical interest. Firstly, they are fully tractable via state-of-the-art shell-model calculations. Specifically, the \emph{SDPF-U} interaction \cite{PhysRevC.79.014310} was designed to describe the physics inside the $N$ = 28 island of inversion and has succeeded in reproducing excitation spectra in the high-$Z$ part of this region \cite{PhysRevC.81.064329}. The merging of the $N$ = 28 and $N$ = 20 islands of inversion is well described by the \emph{SDPF-U} Mix interaction \cite{Caurier2014}, even though the predictions of the two interactions significantly differ in lighter isotopes \cite{PhysRevLett.122.052501,PhysRevLett.122.222501}. 

Open-shell medium-mass nuclei also provide important benchmarks for rapidly developing nuclear \emph{ab initio} methods and modern theories of nuclear interactions based on chiral effective field theory.
In this context argon isotopes offer a complementary picture to the calcium chain that constitutes a traditional testing ground. 
One such approach, the valence-space formulation of the In-Medium Similarity Renormalization Group (VS-IMSRG) \cite{Tsuk12SM,Bogn14SM,Stro16TNO,Stro17ENO,Stro19ARNPS}, opened \emph{ab initio} theories to essentially all nuclei accessible to the nuclear shell model, including fully open-shell exotic systems. The
VS-IMSRG provides an adequate description of the emergence of the $N$ = 32 and $N$ = 34 sub-shell closures around the calcium chain \cite{Michimasa2018,PhysRevC.99.064303,PhysRevLett.120.062503,Moug18Cr}, but its ability to simultaneously describe the collapse of the $N$ = 28 closure has not yet been tested. Another approach, the self-consistent Green's function formalism in its Gorkov formulation (SCGF) \cite{PhysRevC.84.064317}, can now target open-shell nuclei and thus allows the testing of various Hamiltonians along complete isotopic chains \cite{Lapoux16, som2019}.

In this article we report on the high-precision measurement of the neutron rich $^{46-48}$Ar isotopes. The question of the persistence of the $N$ = 28 gap is revisited in light of the new high-precision data. The new binding energy trends are first compared to predictions from the \emph{SDPF-U} shell-model interaction, which is believed to well describe physics in this region of deformation. We then extend our theoretical investigations to VS-IMSRG calculations, to provide a first test with respect to the evolution of the $N=28$ shell closure below calcium. 
Finally, we present results from SCGF calculations of open-shell isotopes around the calcium chain using the recently derived NN+3N(lnl) chiral Hamiltonian~\cite{som2019}.

\section{Experiment}

\begin{figure*}
\centering
\includegraphics[scale=1]{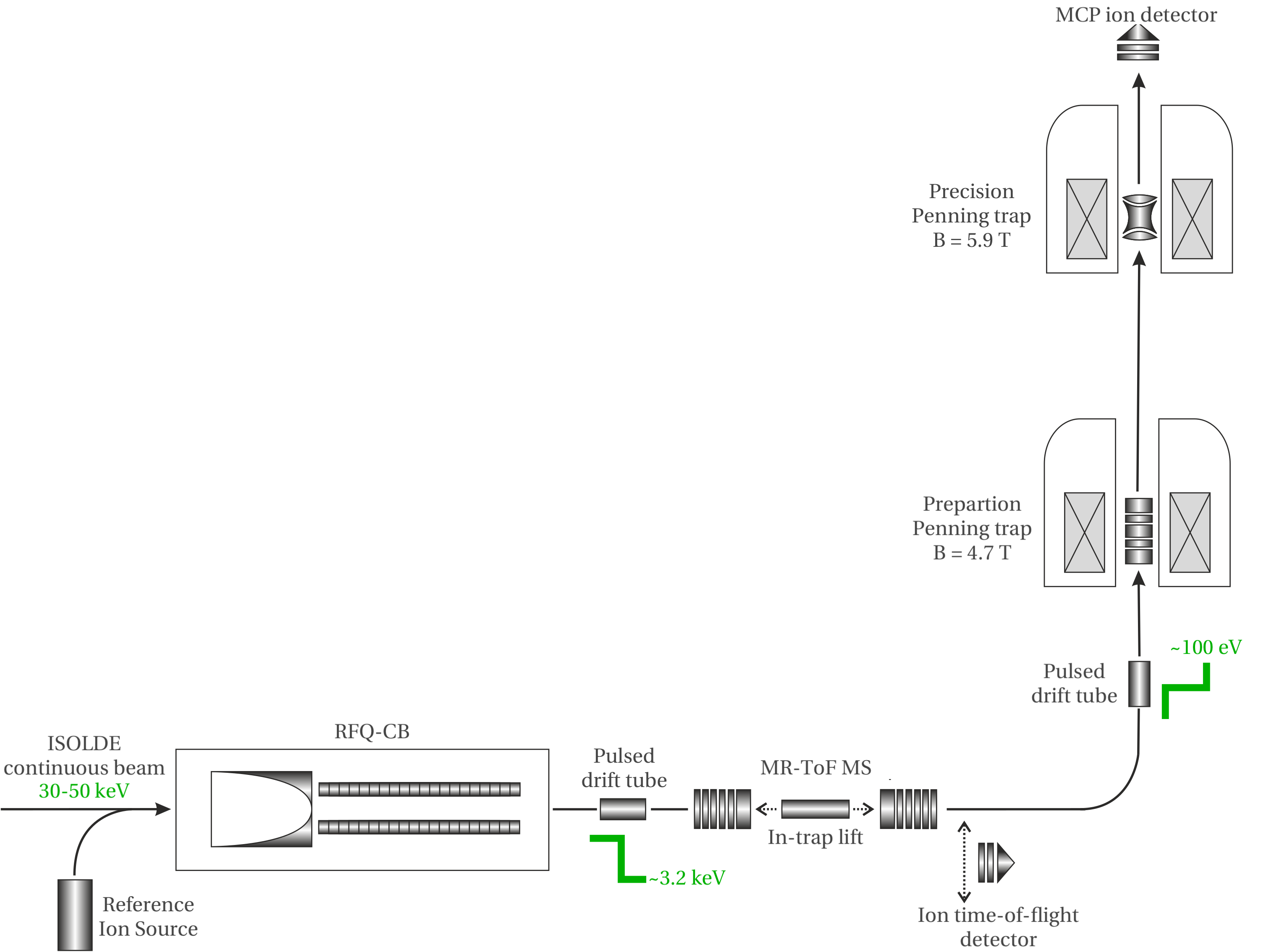}
 \caption{Schematic representation of the ISOLTRAP on-line mass spectrometer. The typical kinetic energy of the ions at various stages of the ISOLTRAP apparatus is shown in green. For details see \cite{Mukherjee-EurPhysJA,Kreim-NuclInstrumMethodsB.317.492}.}
\label{isoltrap_sketch}
\end{figure*}

The measurements reported in this article were performed at the radioactive ion-beam facility ISOLDE at CERN \cite{ISOLDE_2017} in July 2015 and August 2017. In both experiments, the radioisotopes of interest were produced using a thick UC$_{x}$ target which was bombarded with a primary beam of 1.4-GeV protons delivered by the PS-Booster. A VADIS VD7 plasma-ion source was used for ionization. This source was equipped with a water-cooled tantalum transfer line which inhibits the effusion of the less volatile species towards the active volume of the source \cite{doi:10.1063/1.3271245}. The obtained flux of ions was accelerated to a kinetic energy of 30/50 keV in 2015/2017, respectively. Prior to its delivery to the ISOLTRAP on-line mass spectrometer, the isobars of interest were selected using the ISOLDE High-Resolution (magnetic-dipole) Separator (HRS).

A schematic representation of the ISOLTRAP mass-spectrometer \cite{Mukherjee-EurPhysJA,Kreim-NuclInstrumMethodsB.317.492} is shown in Fig. \ref{isoltrap_sketch}. The radioactive ions were first accumulated in a linear radio-frequency cooler-buncher trap (RFQ-CB) \cite{Herfurth2001254},
where the emittance of the incoming beam was reduced in a few milliseconds through collisions with the helium buffer gas (see Table \ref{ArTable} for details).

The ions were extracted from the RFQ-CB in short bunches, were decelerated by a pulsed drift cavity to a kinetic energy of $\approx$ 3.2 keV and then injected into a Multi-Reflection Time-of-Flight Mass Separator (MR-ToF MS) \cite{WOLF201282,WOLF2013123}. There, the bunch of ions was reflected back and forth repeatedly between two electrostatic mirrors. As a result, the various isobaric species constituting the ISOLDE beam were separated in flight time. The beam composition was studied by measuring the time of arrival of the different beam constituents to a secondary electron multiplier placed behind the MR-ToF MS. The experimental duty cycle was adapted according to the nature and abundance of the contamination (see Table \ref{ArTable} for details). Typically, the beam was kept for 1000 revolutions inside the MR-ToF, corresponding to a trapping time of $\approx$16 ms. In all cases, the radioactive species were unambiguously identified by observing the effect with and without proton beam. After separation, the selection of the species of interest was achieved by optimising the timing and length of the extraction pulse from the MR-ToF MS \cite{WIENHOLTZ2017285}.

Being a noble gas, argon is characterized by a large first ionization potential and thus is prone to charge-exchange reactions with neutral impurities contained in the helium gas of the RFQ-CB \cite{DELAHAYE2004604}. The charge-exchange half-life inside the RFQ-CB was determined by monitoring the evolution of the number of stable argon isotopes behind the MRToF-MS as a function of the RFQ-CB cooling time. During the 2017 experiment, stable $^{38}$Ar$^{+}$ was used and the charge-exchange half-life was determined to initially be 23(2) ms. In order to purify the buffer gas, the He injection line was immersed in a bath of liquid nitrogen. Six hours after the installation of this cold trap, the charge-exchange half-life had improved to 50(13) ms. In 2015, the buncher charge-exchange half-life with the cold trap was estimated to be 33(5) ms for $^{36}$Ar$^{+}$. In both runs, the charge-exchange phenomenon was exploited to distinguish the argon isotopes from the contaminants by monitoring the count-rate loss in the argon time-of-flight window as a function of the RFQ-CB trapping time.

\begin{table*}
\centering
 \caption{Summary of the production, preparation and measurement conditions for the isotopes $^{46-48}$Ar. For the ToF-ICR data, the exact quadrupole-excitation time applied in the measurement Penning trap is given. For the Ramsey-type ToF-ICR resonances, the total quadrupole excitation time is presented as $\tau^{RF}_{on}$-$\tau^{RF}_{off}$-$\tau^{RF}_{on}$.} 
\label{ArTable}
\begin{ruledtabular}
  \begin{center}
    \begin{tabular}{c c c c c | c c c c c c}
        \multicolumn{5}{c|}{\textbf{Production}} & \multicolumn{5}{c}{\textbf{Preparation/Measurement}} \\ \cline{1-11}
        
        Date & Target/Line & Source & Sep. & Energy & Ion & RFQ-CB & MR-ToF MS & Prep. Trap & Meas. Trap & Method \\ \cline{1-11}
        
        \multirow{4}{*}{July 2015} & \multirow{4}{*}{UC$_{x}$/Ta} & \multirow{4}{*}{VD7} & \multirow{4}{*}{HRS} & \multirow{4}{*}{30~kV} &  $^{46}$Ar$^{+}$ & 10~ms & 16.3~ms & 104~ms & 200~ms & 2 $\times$ ToF-ICR\\  \cline{6-11}
       
       & & & & & \multirow{3}{*}{$^{47}$Ar$^{+}$} &  \multirow{3}{*}{15~ms} & \multirow{3}{*}{19.8~ms} & \multirow{3}{*}{104~ms} & 100~ms & 2 $\times$ ToF-ICR \\
       
       & & & & & & & & & 200~ms & 1 $\times$ ToF-ICR \\
       
       & & & & & & & & & 10-80-10~ms & 2 $\times$ Ramsey ToF-ICR \\ \cline{1-11}
       
       Aug. 2017 & UC$_{x}$/Ta & VD7 & HRS & 50~kV & $^{48}$Ar$^{+}$ & 5~ms & 16.7~ms & & & 95 $\times$ 1000revs MR-ToF \\ 
        
    \end{tabular}
   \end{center}
  \end{ruledtabular}
\end{table*}

After a 90-degree bend, the purified ion beam entered ISOLTRAP's vertical transport section and was captured in the preparation Penning trap \cite{RAIMBAULTHARTMANN1997378}. In this He-filled device, further beam purification was achieved using the so-called mass-selective resonant buffer-gas cooling technique \cite{SAVARD1991247}. Once again, a cold trap was used to purify the He-gas injection line. After  installation of the cold traps, the charge exchange half-life in the preparation Penning trap was 223(38) ms. Consequently, a rather short processing time (see Table \ref{ArTable} for details) was used. Finally, the ion bunch was transported to the precision Penning trap, where the free cyclotron frequency of the ion of interest was measured using the Time-of-Flight Ion-Cyclotron-Resonance (ToF-ICR) technique \cite{Koenig-IntJMassSpectrom.142.95}.

The ion mass $m_{\textit{ion,x}}$ is connected to its cyclotron frequency by the relation:
\begin{equation}
\nu_{c,x} = \frac{q_{x} B}{2 \pi m_{\textit{ion,x}}},
\end{equation}
where $q_{x}$ is the ion's charge (in the following we consider $q_{x}$ = \emph{e} for all species) and $B$ is the strength of the confining magnetic field. The calibration of the magnetic field is performed by measuring the cyclotron frequency $\nu_{\textit{c,ref}}$ of a reference species of well-known mass $m_{\textit{ion,ref}}$ shortly before and after the measurement of the species of interest. The cyclotron frequency of the reference species is then linearly interpolated to the time at which the measurement of the ion of interest was performed. From the experimentally measured cyclotron-frequency ratio:
\begin{equation}
r_{\textit{ref},x} = \frac{\nu_{c,\textit{ref}}}{\nu_{c,x}} = \frac{m_{ion,x}}{m_{ion,\textit{ref}}},
\end{equation}
the atomic mass of the species of interest is calculated according to the relation: 
\begin{equation}
m_{\textit{atom,x}} = r_{\textit{ref},x }(m_{\textit{atom,ref}} - m_{e}) + m_{e},
\end{equation}
where $m_{e}$ is the electron mass \cite{Sturm2014}.

Sometimes the low yield and/or short half-life of an ion species make a Penning-trap measurement impossible. In this case, the MR-ToF MS can be used as a mass spectrometer in its own right. The relationship between an ion mass-over-charge ratio $\frac{m_{\textit{ion,x}}}{q_{x}}$ and its time-of-flight $t_{x}$ is given by \cite{Guilhaus1995}:
\begin{equation}
t_{x} = a \sqrt{\frac{m_{ion,x}}{q_{x}}} + b,
\end{equation}
where $a$ and $b$ are calibration parameters which can be determined by measuring the flight times $t_{1,2}$ of two reference ions with well-known masses $m_{1,2}$ and charges $q_{1,2}$. The mass of an ion is calculated from the relation \cite{Wienholtz-Nature.498.346}:

\begin{align} 
\sqrt{\frac{m_{ion,x}}{q_{x}}} & = C_{\textit{TOF}} \left( \sqrt{\frac{m_{ion,1}}{q_{1}}}-\sqrt{\frac{m_{ion,2}}{q_{2}}} \right) \nonumber \\
 & + \frac{1}{2} \left( \sqrt{\frac{m_{ion,1}}{q_{1}}}+\sqrt{\frac{m_{ion,2}}{q_{2}}} \right),
\end{align}

with:
\begin{equation}
C_{\textit{TOF}} = \frac{2t_{x}-t_{1}-t_{2}}{2(t_{1}-t_{2})}.
\end{equation}

\subsection{The $^{46}$Ar mass}

During the 2015 experiment, although significant amounts of $^{92}$Kr$^{2+}$ were present in the beam, the most detrimental $A$ = 46 contaminant was the stable $^{34}$S$^{12}$C$^{+}$ molecular ion. A mass resolving power of $R =\frac{m}{\Delta m} =$ 2 $\times$ 10$^{5}$ is needed to separate $^{46}$Ar$^{+}$ from this contaminant. As a result, a mixture of the two species was transported to the measurement Penning trap, where a ratio of 3:1 in favor of the contaminant species was initially observed. Fortunately, after a few days the outgasing of the $^{34}$S$^{12}$C$^{+}$ molecular ion from the target unit reversed this ratio. 

To enhance the collection of argon ions even further, the ISOLTRAP cycle was synchronized to the proton impact on the ISOLDE target and delayed by 50~ms to accumulate the argon ions at the maximum of their release from the target. The RFQ-CB cooling time was also reduced from 
25~ms to 10~ms to minimise charge-exchange losses. These modifications meant that two quasi-pure ToF-ICR resonances of $^{46}$Ar$^{+}$ were recorded. A quadrupole-excitation time of 200~ms was used in both cases (see Table \ref{ArTable} for details). 

\begin{figure}
\centering
\includegraphics[scale=0.35]{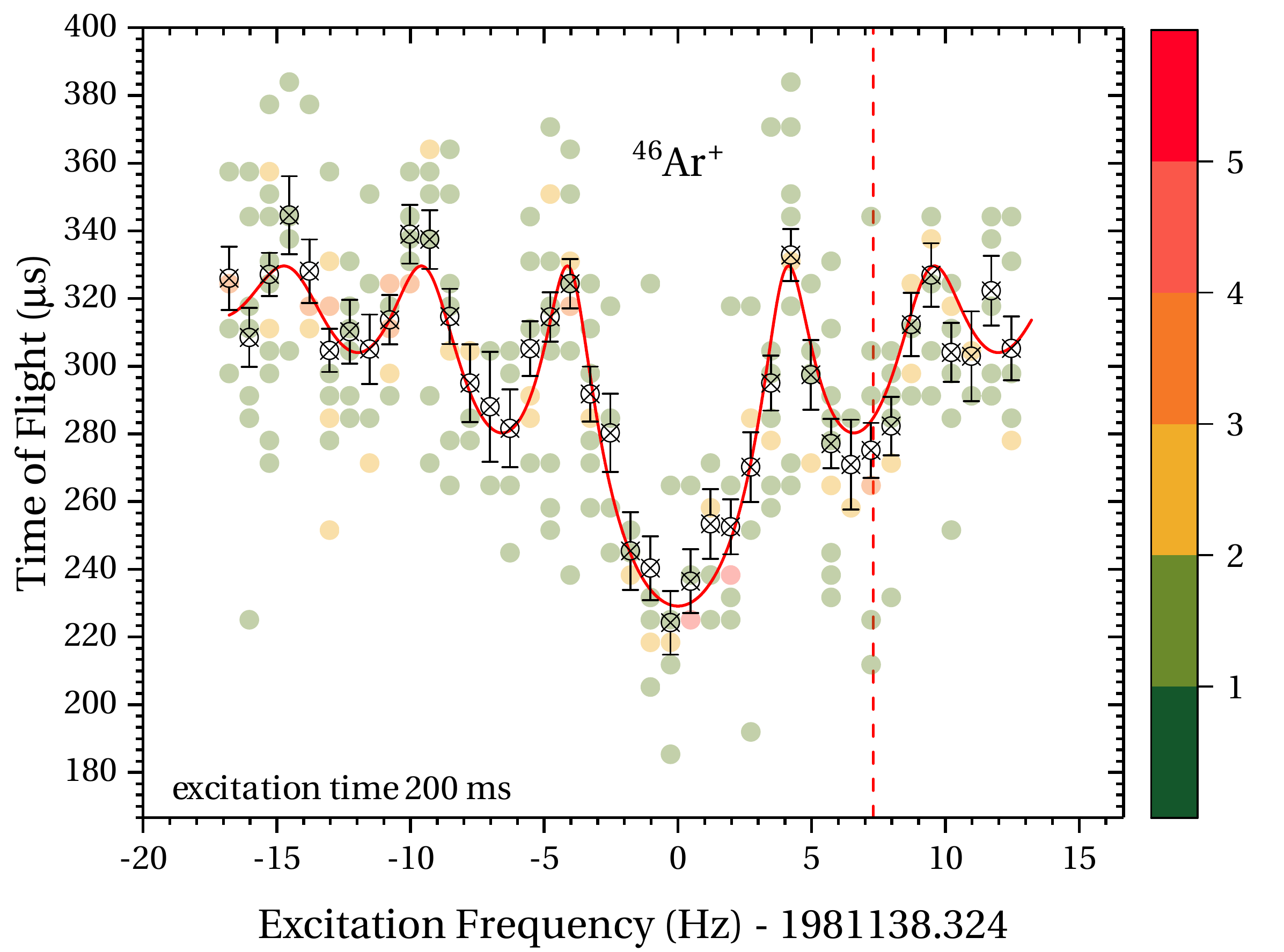}
\caption{A typical one-pulse ToF-ICR resonance \cite{Koenig-IntJMassSpectrom.142.95} of $^{46}$Ar$^{+}$ ($\tau_{on}^{RF}$ = 200 ms). The color-map represents the ion events recorded in each (frequency;tof) bin. The mean and standard deviation of the time-of-flight distribution recorded for each frequency is shown as open circles while the red line shows the result of the least-squares adjustment of the theoretical line shape to these data points. The vertical dashed line indicates the expected cyclotron frequency of the contaminant species $^{34}$S$^{12}$C$^{+}$.}
\label{46Ar_toficr}
\end{figure}

Because of the presence of $^{34}$S$^{12}$C$^{+}$, extra care was taken during the analysis procedure. In the present case, a vast majority of ejections out of the measurement Penning trap resulted in no ions detected (average count rate of 0.2 ions/ejections) while 250 events were recorded with only one ion detected. This number drops by a factor 5 for two ions detected per ejection and even more significantly for three ions or more. As a result, the so-called z-class analysis, a procedure described in \cite{kellerbauer2003} to estimate the effect of contaminants in ToF-ICR resonances could not be performed here. To limit the impact of residual contamination, the analysis was performed using the events where only one ion was detected after the measurement trap.

\begin{figure}
\centering
\includegraphics[scale=0.35]{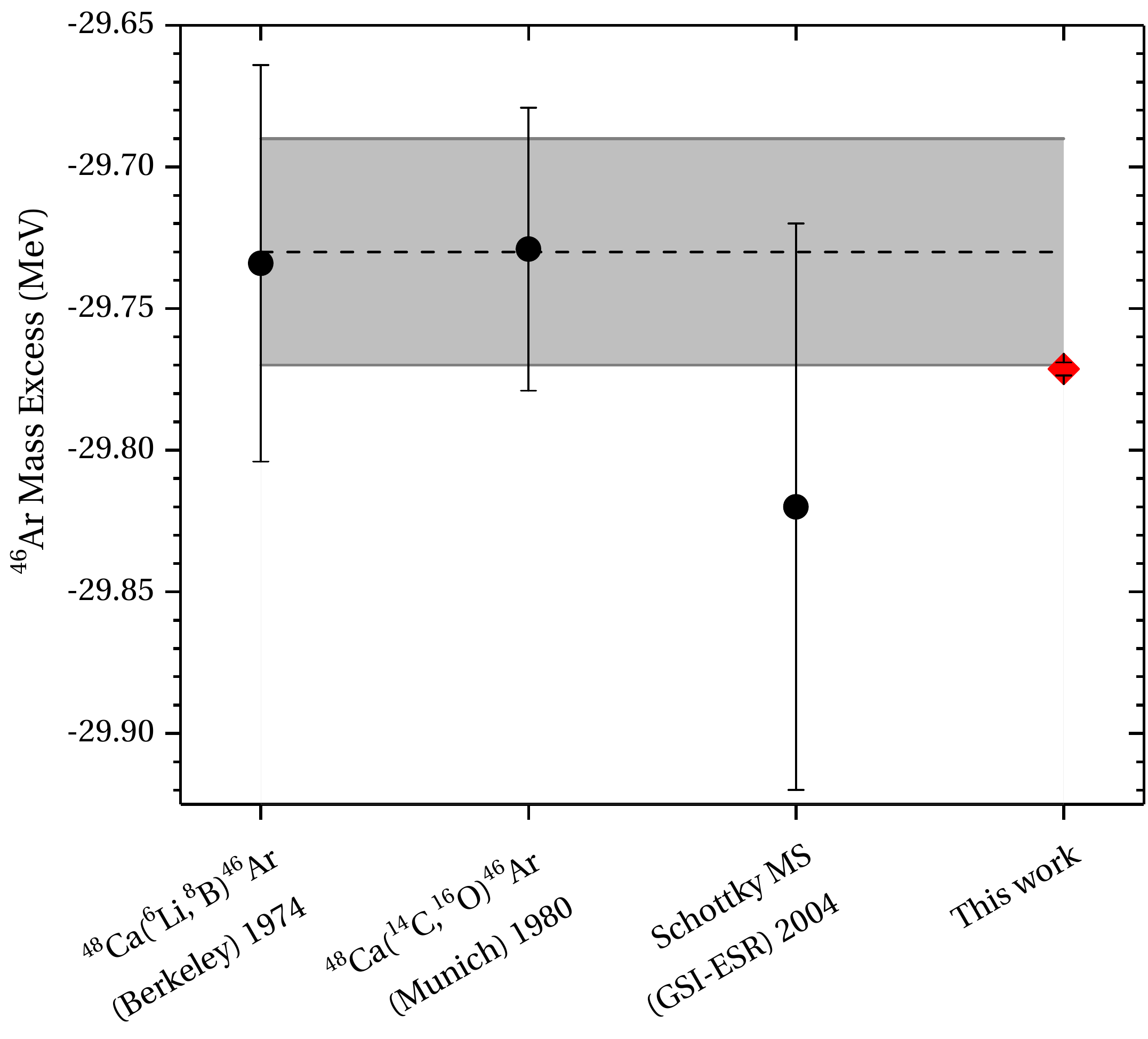}
\caption{Comparison between the value for the $^{46}$Ar mass excess obtained in this work (red diamond) and the ones obtained in previous works \cite{PhysRevC.9.2067,PhysRevC.22.2449,Matos_thesis}. The black dashed line marks the AME2012 value while the grey band represents the AME2012 one standard deviation \cite{AME2012}. For the red diamond point, the uncertainty is smaller than the size of the point.}
\label{46arcomp}
\end{figure}

A typical resonance is shown in Fig. \ref{46Ar_toficr}. The purity of the resonance is attested by two factors. First, around the free cyclotron frequency of $^{34}$S$^{12}$C$^{+}$ (indicated by the vertical red dashed line 
in Fig. \ref{46Ar_toficr}) very few ion counts are present between 220 and 240~\textmu s, meaning that very few excited contaminant ions were recorded. Second, close to zero frequency detuning, the time-of-flight distribution for each frequency value does not exhibit a significant amount of ion events detected at time-of-flights around 330~\textmu s, which would indicate the presence of unexcited contaminant ions. 

In the present case, $^{39}$K$^{+}$ (atomic mass $m_{^{\text{39}}\text{K}}$ = 38963706.487(5)~\textmu u \cite{AME2016}) was used as a reference for the magnetic-field calibration. Taking into account the various sources of systematic uncertainties described in \cite{kellerbauer2003}, one obtains the mean frequency ratio in Table \ref{Ar_results}. This translates to an atomic mass excess of ME($^{46}$Ar) = -29771.3(23)~keV. Fig. \ref{46arcomp} shows a comparison between the value from this work and that obtained from previous measurements. When compared to the AME2012 value \cite{AME2012}, our new measurement deviates by 41.3~keV but is 20 times more precise. The AME2012 value was primarily determined through two \emph{Q}-value measurements: one in 1974 using the $^{48}$Ca($^{6}$Li,$^{8}$B)$^{46}$Ar reaction \cite{PhysRevC.9.2067} and another in 1980 using the $^{48}$Ca($^{14}$C,$^{16}$O)$^{46}$Ar reaction \cite{PhysRevC.22.2449}. These results agree with the new mass and were complemented by a measurement performed using the Isochronous Mass Spectrometry technique at the FSR-ESR storage ring (GSI, Germany) \cite{Matos_thesis} in 2004, which also agrees but had no weight in the evaluation due to the larger uncertainty.

\begin{table*}[!ht]
\centering
 \caption{Final frequency ratios ($r_{ref,x} = \nu_{c,ref} / \nu_{c,x}$), time-of-flight ratios ($C_{ToF}$) and mass excesses of the argon isotopes measured in this work. Values of the mass excesses from the Atomic-Mass Evaluation 2016 (AME2016) \cite{AME2016} are given for comparison. Values from AME2012 are also given \cite{AME2012} (\# designates AME2012 extrapolated value).The masses of the reference ions were also taken from AME2016. Experimental half-lives are taken from the NUBASE2016 evaluation \cite{Nubase2016}.} 
\begin{ruledtabular}
  \begin{center}
    \begin{tabular}{c c c c c c c}
    & & & & \multicolumn{3}{c}{\textbf{Mass Excess (keV)}} \\ \cline{5-7} 
    \textbf{Species} & \textbf{Half-life} & \textbf{Reference} & \textbf{ ratio \emph{r} or $C_{ToF}$} & \textbf{This work} & \textbf{AME2016} & \textbf{AME2012} \\ \cline{1-7}
    $^{46}$Ar & 8.4(8) s & $^{39}$K & $r_{ref,x}$ = 1.1797680972(640) & -29771.3(23) & -29772.9(11) & -29730(40)\\  \cline{1-7}
    $^{47}$Ar & 1.23(3) s & $^{39}$K & $r_{ref,x}$ = 1.2055547092(340)  & -25367.3(12) & -25366.3(11) & -25210(90)\\  \cline{1-7}
    $^{48}$Ar & 415(15) ms & $^{32}$S$^{16}$O/$^{85}$Rb & $C_{ToF}$ = 0.499715668(560)  & -22355(17) & -22280(310) & -22440\# (300)\# \\
        \end{tabular}
   \end{center}
  \end{ruledtabular}
   \label{Ar_results}
\end{table*}

\subsection{The $^{47}$Ar mass}

\begin{figure}
  \centering 
\includegraphics[scale=0.35]{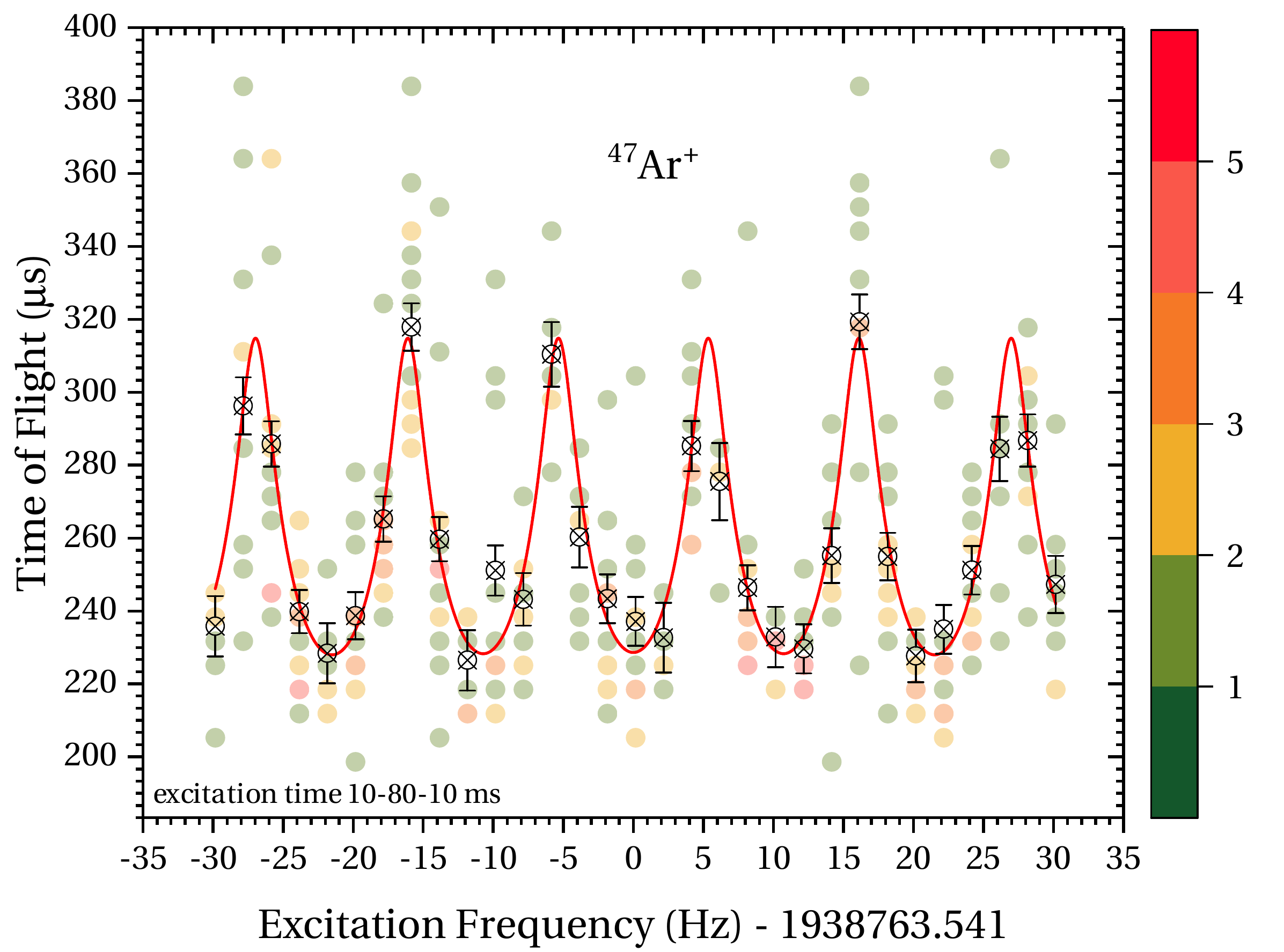}
 \caption{A typical ToF-ICR resonance of $^{47}$Ar$^{+}$ using the Ramsey-type excitation scheme ($\tau_{on}^{RF}-\tau_{off}^{RF}-\tau_{on}^{RF}$ = 10~ms - 80~ms - 10~ms) \cite{George-IntJMassSpectrom.264.110,PhysRevLett.98.162501}. The color-map represents the ion events recorded in each (frequency;tof) bin. The mean and standard deviation of the time-of-flight distribution recorded for each frequency value is shown as open circles while the red line shows the result of the least-squares adjustment of these data points to the theoretical line shape.}
\label{47Ar_ramsey}
\end{figure}

The $^{47}$Ar$^{+}$ ions were well separated from the other contaminants so that a pure beam was transported to the measurement Penning trap. The details of the ISOLTRAP measurement cycle are summarized in Table \ref{ArTable}. In total, three ToF-ICR resonances were recorded using a quadrupole-excitation time of 100 ms and 200 ms. In addition, two ToF-ICR resonances in the Ramsey-type excitation scheme \cite{George-IntJMassSpectrom.264.110,PhysRevLett.98.162501} were recorded. This excitation scheme is characterized by the application of two short radio-frequency pulses of duration $\tau_{on}^{RF}$ which are coherent in phase and separated by a waiting time $\tau_{off}^{RF}$. For the same total excitation time, this method offers a three-fold precision improvement in the determination of the free cyclotron frequency of an ion when compared to the single-pulse ToF-ICR method. In the present case, a $\tau_{on}^{RF} - \tau_{off}^{RF} - \tau_{on}^{RF}$ = 10~ms - 80~ms - 10~ms excitation scheme was used. A typical example of such a Ramsey-resonance is shown in Fig. \ref{47Ar_ramsey}.

\begin{figure}
\centering
\includegraphics[scale=0.35]{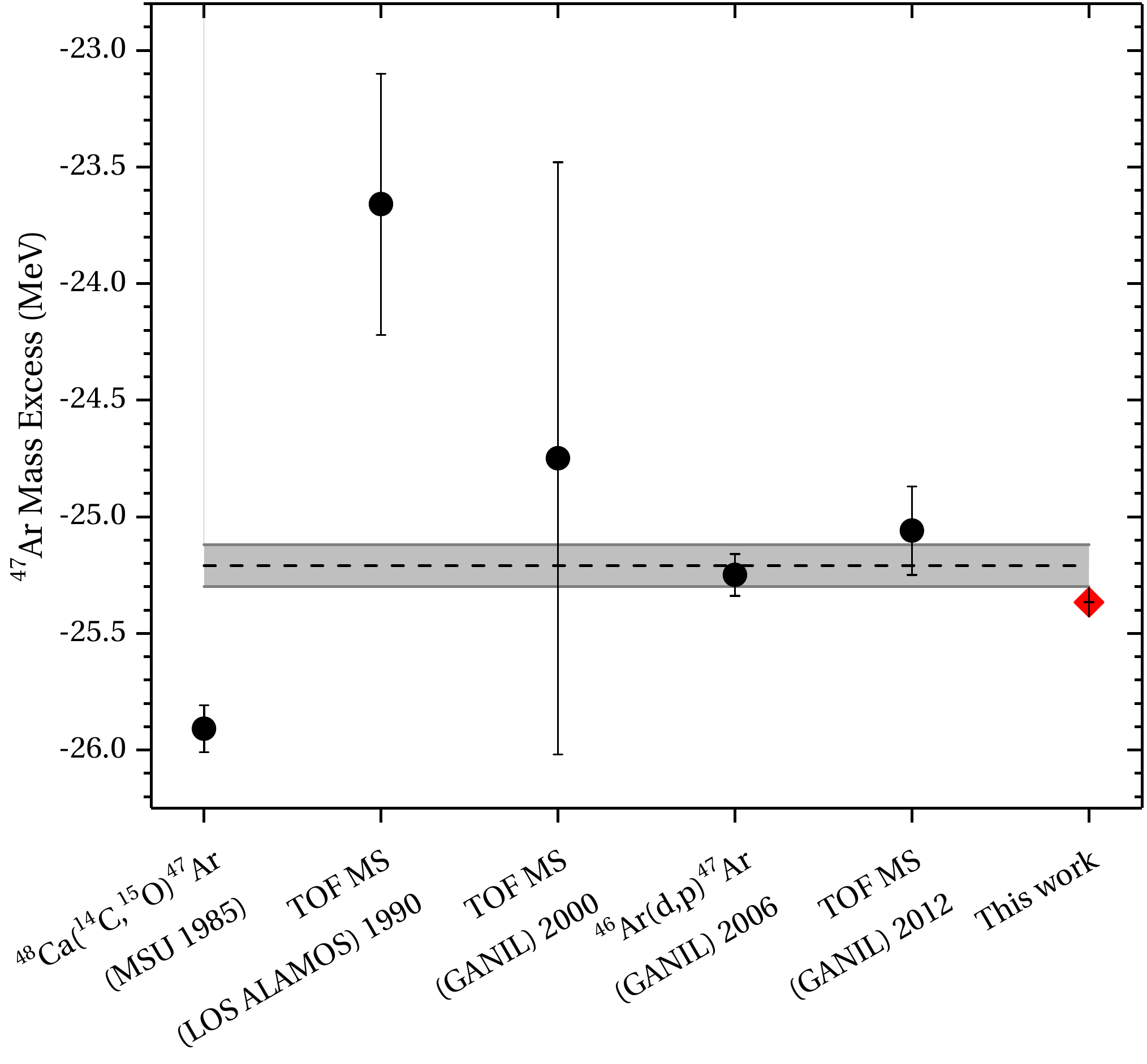}
\caption{Comparison between the value for the $^{47}$
Ar mass excess obtained in this work (red diamond) and the ones obtained in previous works \cite{BENENSON198587,Tu1990,PhysRevLett.84.5062,Gaudefroy2006,Gaudefroy2012}. The black dashed line marks the AME2012 value while the grey band represents the AME2012 one standard deviation \cite{AME2012}. For the red diamond point, the uncertainty is smaller than the size of the point.}
\label{47arcomp}
\end{figure}

Here, $^{39}$K$^{+}$ ions were also used for the calibration of the magnetic field. The mean frequency ratio of Table \ref{Ar_results} can be used to derive an atomic mass excess value  ME($^{47}$Ar) = -25367.3(12) ~keV. Figure \ref{47arcomp} shows the comparison between the new value from this work and previous measurements. Compared to the AME2012 value, our measurement provides a $\sim $ 90-fold improvement in precision and is 157~keV more bound. The AME2012 \cite{AME2012} value is mainly influenced by a measurement of the \emph{Q}-value of the reaction $^{46}$Ar(d,p)$^{47}$Ar \cite{Gaudefroy2006}. In this study the authors reported a 700-keV deviation to a previous measurement obtained from the reaction $^{48}$Ca($^{14}$C,$^{15}$O)$^{47}$Ar \cite{BENENSON198587}. In addition, the AME2012 also includes two time-of-flight measurements of $^{47}$Ar \cite{Tu1990,PhysRevLett.84.5062} which due to their large uncertainty bore no significant weights in the evaluation. The close proximity between the mass excesses of $^{46-47}$Ar reported in this work and that tabulated in the AME2016 \cite{AME2016} is due to the fact that a very preliminary version of the results presented in this work was communicated to the AME evaluators. Apart from this preliminary value, the AME2016 \cite{AME2016} also includes a time-of-flight measurement performed at GANIL \cite{Gaudefroy2012}. As shown in Table \ref{Ar_results} our results dominate the weight in the final AME2016 adjustment.

\subsection{The $^{48}$Ar mass}

\begin{figure*}
\centering
\includegraphics[scale=0.75]{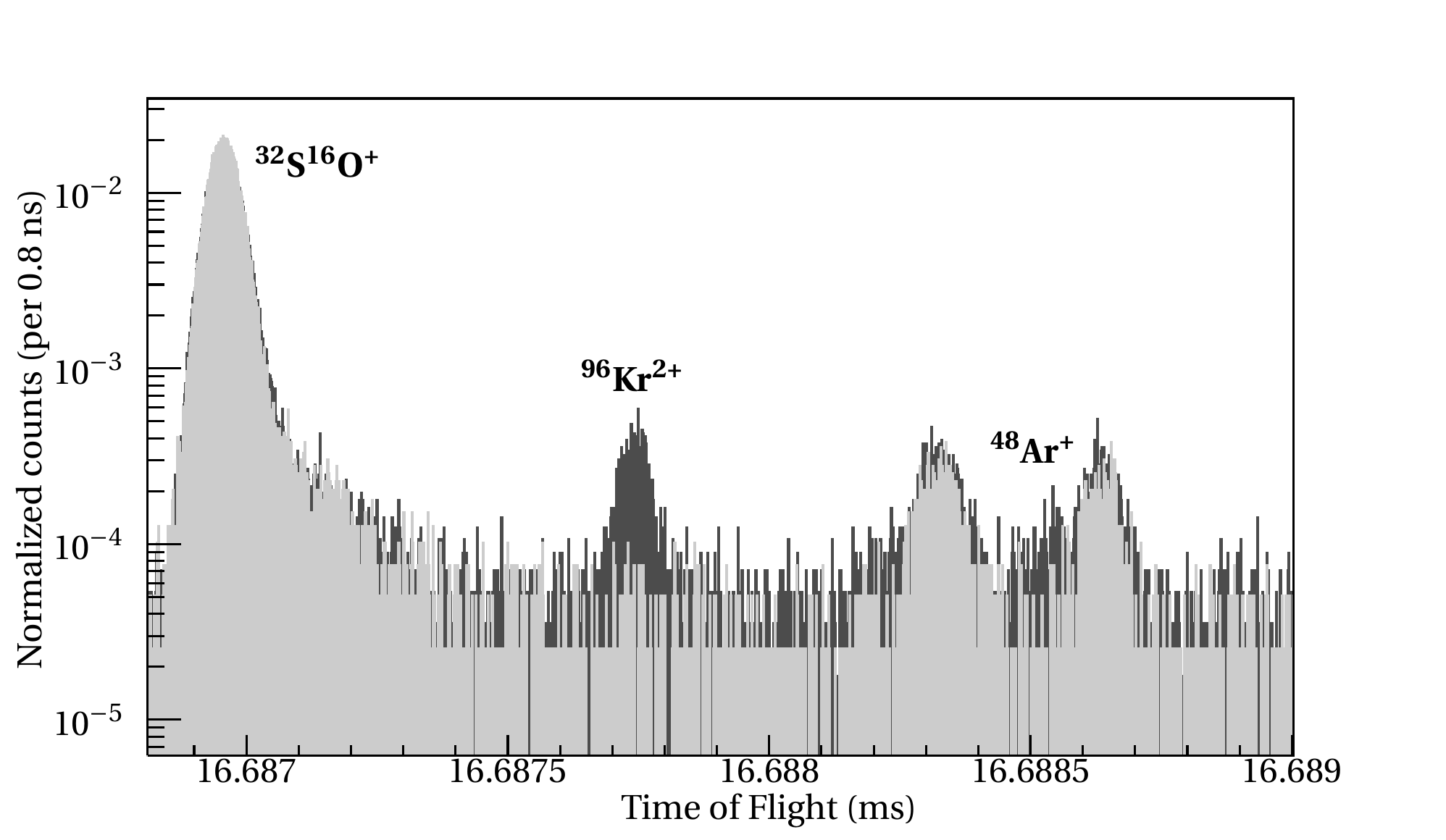}
  \caption{\emph{A} = 48 time-of-flight spectrum after 1000 revolutions inside the MR-ToF MS. The spectrum recorded with protons on target results from the sum of 13 consecutive files and is represented in dark-grey. The light-grey spectrum represents a background measurement performed while the protons were turned off and is the sum of 21 consecutive files.}
  \label{48Ar_MRToF}
\end{figure*}

\begin{figure*}
\centering
\includegraphics[scale=0.75]{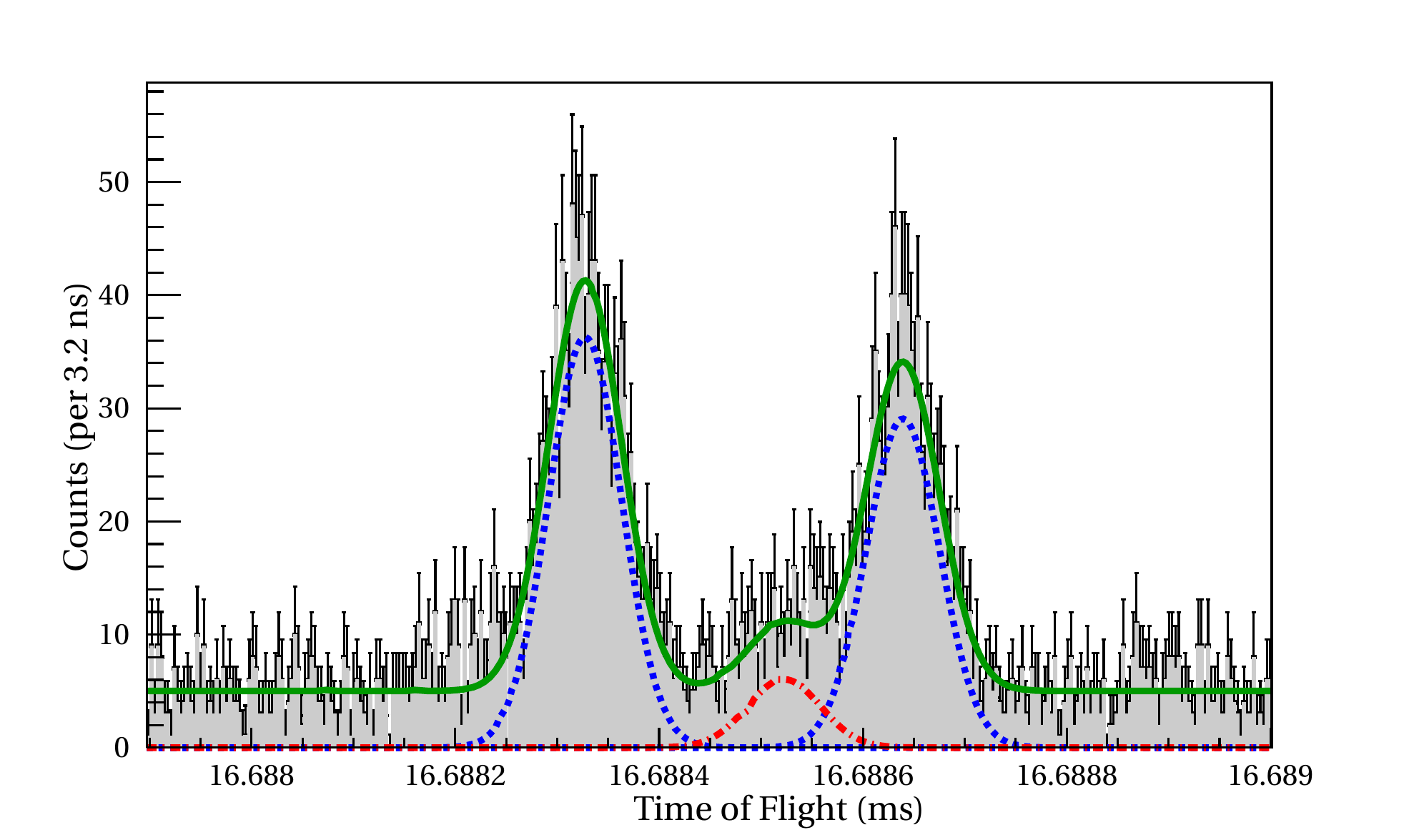}
\caption{The model PDF used to extract the time of flight of $^{48}$Ar$^{+}$. The analysis is performed in a restricted 1.1~\textmu s window. The full PDF is represented as a solid green line while the dashed blue and dashdotted red lines represent the contaminant (two Gaussians) and signal (one Gaussian) components, respectively.}
\label{48Ar_models}
\end{figure*}

The previous measurement campaign was followed in 2017 by an experiment targeting the measurement of $^{48}$Ar$^{}$. In order to establish the presence of the radioactive $^{48}$Ar$^{+}$ isotope in the ISOLDE beam, a reference time-of-flight histogram was built from 21 consecutive files recorded with the MR-ToF MS without protons on target. This histogram was compared with a histogram resulting from the sum of 13 consecutive files recorded with protons on target. To allow the comparison between the two spectra both were normalized to their total number of recorded events and superimposed. As seen from Fig. \ref{48Ar_MRToF}, the $A$ = 48 ISOLDE beam was found to be dominated by the presence of the stable $^{32}$S$^{16}$O$^{+}$ molecular ion which was unambiguously identified by measuring its cyclotron frequency in ISOLTRAP's measurement Penning trap. At later TOF, a double-peak structure corresponding to stable contamination is also visible. The yields of these species were too low to allow for the determination of their cyclotron frequencies using the measurement Penning trap. Their times of flight were compared to a wide variety of singly- and doubly-charged atomic and simple molecular species, none matched. 

With protons on target, a $^{96}$Kr$^{2+}$ peak became clearly visible. Synchronizing the start of the experimental cycle with the proton impact on target, an excess of counts also appeared between the two stable undetermined species within the expected time-of-flight window for $^{48}$Ar$^{+}$. Varying the RFQ-CB cooling time from 20 to 150 ms, the absolute strength of this signal was extracted using the binned, extended maximum likelihood estimation method within a restricted time-of-flight window of 1.1~\textmu s \cite{doi:10.1002/9783527653416.ch2}. The probability-density function (PDF) of the fit was composed of the sum of two Gaussian PDFs (describing the two stable contaminants) and a uniform component (to capture the rather high level of baseline background) while the signal component was also considered to be described by a Gaussian PDF. In addition, the three Gaussian PDFs were assumed to share the same width parameter. In total eight parameters were left free during the estimation. Hence, we found that the strength of the studied signal decreases when the RFQ-CB cooling time is increased at a rate consistent with the observed charge-exchange half-life.

In total, eight MR-ToF MS spectra were used to perform the mass determination of $^{48}$Ar$^{+}$. Each of these spectra results from the sum of 8 to 20 individual files recorded consecutively. Within this set of 8 spectra, as few as 30 and as much as 170~ion counts, for a total of 700~ion counts attributed to $^{48}$Ar$^{+}$ were recorded. The same analysis method and parameters as used for estimating the signal strength were kept for the determination of the mean TOF of $^{48}$Ar$^{+}$. 

Figure \ref{48Ar_models} shows a typical example of the adjusted PDF (solid green line). The background component (dashed blue line) and signal components (dash-dotted red line) are also represented. For the $A$ = 48 mass determination, the molecular contaminant $^{32}$S$^{16}$O$^{+}$ present in the $A$ = 48 spectrum (atomic mass $m_{^{\text{32}}\text{S}^{\text{16}}\text{O}}$ = 47966985.794(1)~\textmu u \cite{AME2016}) and $^{85}$Rb$^{+}$ (atomic mass $m_{^{\text{85}}\text{Rb}}$ = 84911789.738(5)~\textmu u \cite{AME2016}) provided by ISOLTRAP's offline ion source (see Fig. \ref{isoltrap_sketch}) were used as references. The obtained mean $C_{ToF}$ parameter and its associated uncertainty can be found in Table \ref{Ar_results}. 

\begin{figure}
\centering
\includegraphics[scale=0.35]{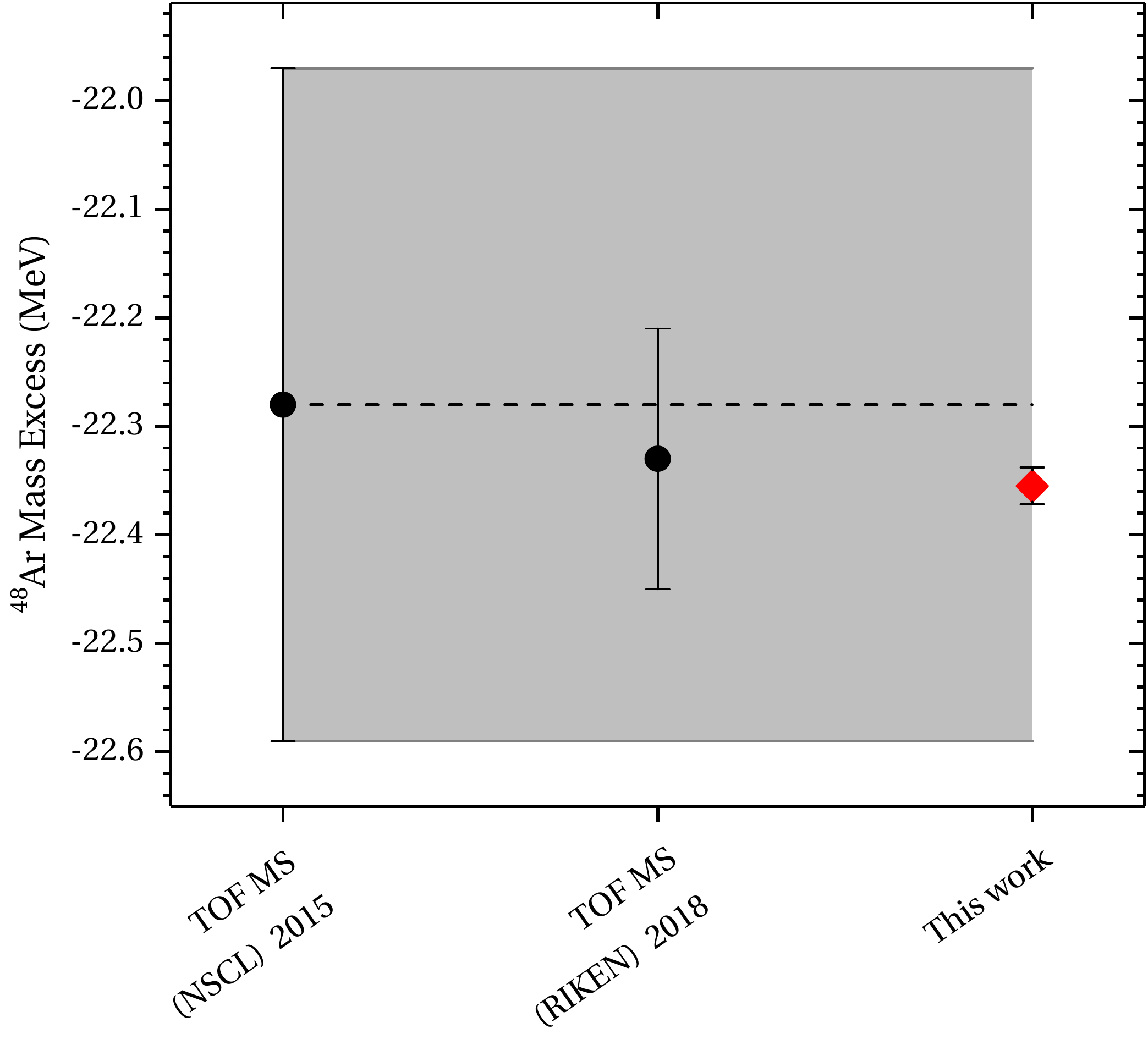}
\caption{Comparison between the value for the $^{48}$Ar mass excess obtained in this work (red diamond) and those obtained in previous works \cite{Meisel2015,Michimasa2018}. The black dashed line marks the AME2016 value while the grey band represents the AME2016 one standard deviation \cite{AME2016}.}
\label{48arcomp}
\end{figure}

When one of the reference species is part of the same time-of-flight spectrum as the ion of interest, the accuracy of the MRToF-MS mass measurement is sensitive to any phenomenon affecting the extracted time-of-flight difference between the two species. In this respect, the main source of systematic uncertainty was found to be the shape of the time-of-flight distributions. As seen in Fig. \ref{48Ar_MRToF}, when sufficient statistics are collected, the time-of-flight peaks exhibit clear tailing towards later flight time. For the sake of consistency, the analysis was performed assuming that all peaks are Gaussian distributed.

To quantify the dependence of the estimated time of flight on the presence of these tails, the time-of-flight estimation was performed a second time for the reference species using the asymmetric peak profile described in \cite{LAN20011}. For each reference species ($i=$ 1 , 2) the time-of-flight differences $\Delta t_{i}$ to the results from the Gausssian PDF were averaged over the 8 spectra yielding the average time-of-flight deviations $\overline{\Delta t_{i}}$. These systematic fit deviations $\overline{\Delta t_{i}}$ were then translated into the individual systematic C$_{ToF}$ uncertainty contributions $\Delta C_{ToF}^{fit,i} = \rvert \frac{\partial C_{ToF}}{\partial t_{i}} \overline{\Delta t_{i} \lvert}$. Finally, all the $\Delta C_{ToF}^{fit,i}$ were added in quadrature to the statistical uncertainty to yield the total $C_{ToF}$ uncertainty. Since the statistics is too low to assess this effect for the $^{48}$Ar$^{+}$ peak, this peak was attributed the same additional uncertainty contribution as that of the isobaric $^{32}$S$^{16}$O$^{+}$ reference, the rest being purely statistical. This effect contributes 35 \% of the final $C_{ToF}$ uncertainty given in Table \ref{Ar_results}. Another systematic-uncertainty source is the so-called peak-coalescence phenomenon \cite{doi:10.1063/1.4796061} whereby the separation between isobaric species is reduced due to their Coulomb interaction. To mitigate this effect the count rate was always kept under $\approx$8 ions/cycle during the measurement, which has been shown from many cross-check measurements to be a safe limit. 

Figure \ref{48arcomp} provides a direct comparison between the new value from the present work and previous measurements. Time-of-flight measurements published in 2015 with the S800 spectrometer at the NSCL \cite{Meisel2015} provided the first mass-excess value for $^{48}$Ar. Very recently, another such measurement was  reported using the SHARAQ spectrometer at RIKEN \cite{Michimasa2018}. This measurement, in agreement with that of NSCL, brought a factor of 2.5 improvement in accuracy. Our measurement of the $^{48}$Ar mass excess (see Table \ref{Ar_results}) shows a factor $\approx$19 improvement in accuracy from the NSCL value while deviating by 74.8~keV. When compared to the RIKEN measurement, the present value provides a factor $\approx$7 improvement in accuracy and deviates by $\approx$25~keV. Both deviations are well within one standard deviation of the respective previous value.

\section{Discussion}

\begin{figure}[!ht]
\centering
\includegraphics[scale=0.35]{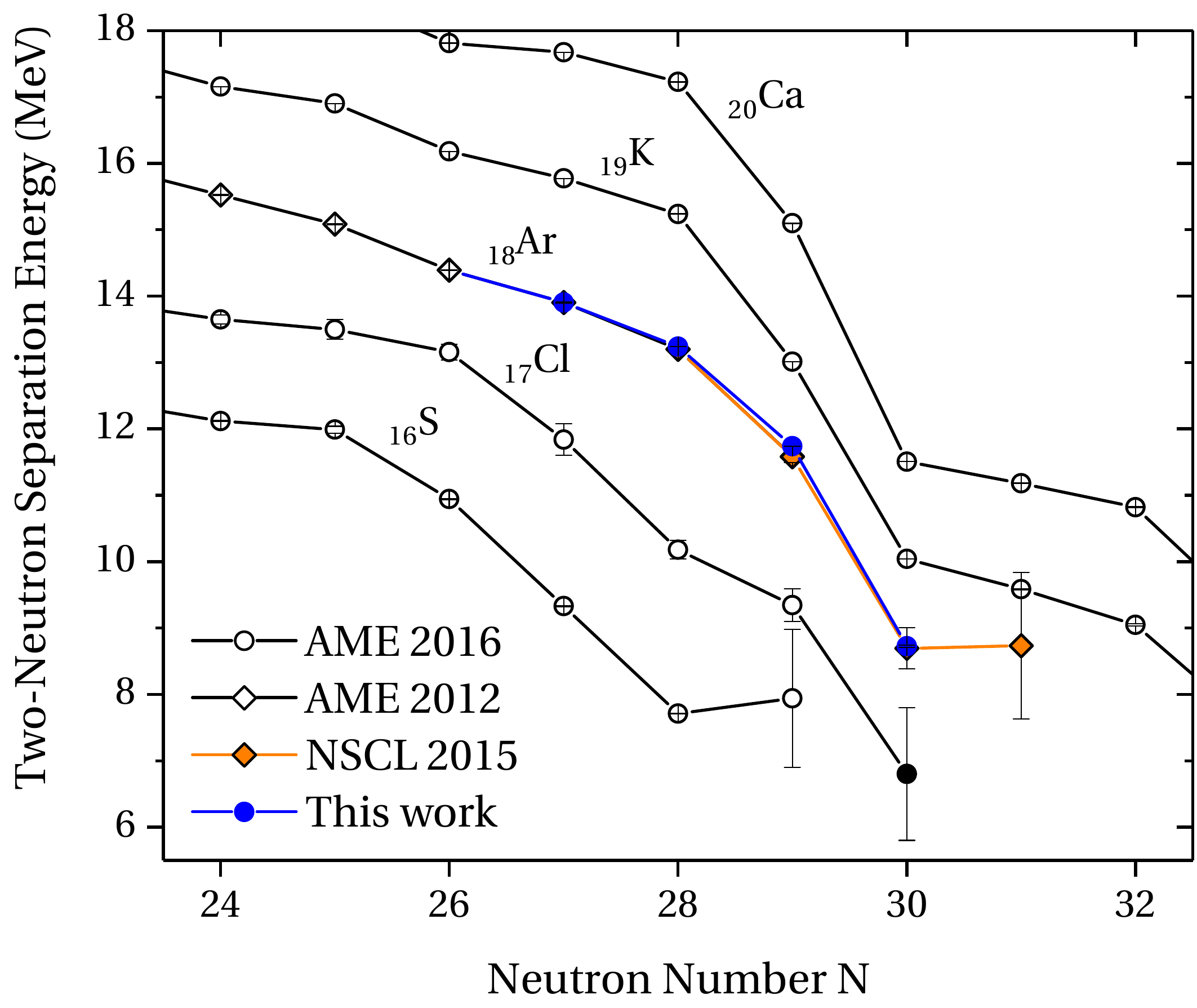}
\caption{Experimental trends of S$_{2n}$ in the $N$=28 region for isotopic chains ranging from sulfur to calcium. For the argon isotopic chain the trend obtained from the AME2012 \cite{AME2012} is represented as open diamonds, the trend extracted from the 2015 NCSL time-of-flight measurements is represented as orange diamonds \cite{Meisel2015} and the trend from this work is shown as blue circles. The values for all the other chains are extracted from the AME2016 mass evaluation \cite{AME2016}. The black circle was obtained using values from \cite{Jurado2007} which are not included in the AME.}
\label{s2n_exp}
\end{figure}

The mass values obtained in this work were used to assess the strength of the empirical \emph{N}~=~28 shell gap for argon. To extract nuclear-structure effects from binding energies, one typically investigates the variation with $N$ or $Z$ of finite binding-energy differences, also called mass filters. One such quantity, the two-neutron separation energy $S_{2n}(N,Z)$, is presented in Fig. \ref{s2n_exp} as a function of $N$ for the isotopic chains with $Z=16-20$. $S_{2n}$ is defined as $ME(N-2,Z)-ME(N,Z)+2M_{n}$ 
where $ME(N,Z)$ represents the mass excess of an isotope with \textit{N} neutrons and \textit{Z} protons and $M_{n}$ is the neutron mass excess. Along an isotopic chain, the $S_{2n}$ values usually follow a steadily decreasing trend,
while at a shell closure, the magnitude of this decrease is markedly larger. Figure \ref{s2n_exp} confirms that the trend of $S_{2n}$ obtained in this work for $Z$~=~18 is not significantly different than the one obtained using the results from \cite{Meisel2015}, from which a strong \emph{N}~=~28 shell-gap in the argon chain was inferred. 

\begin{figure}
\centering
\includegraphics[scale=0.35]{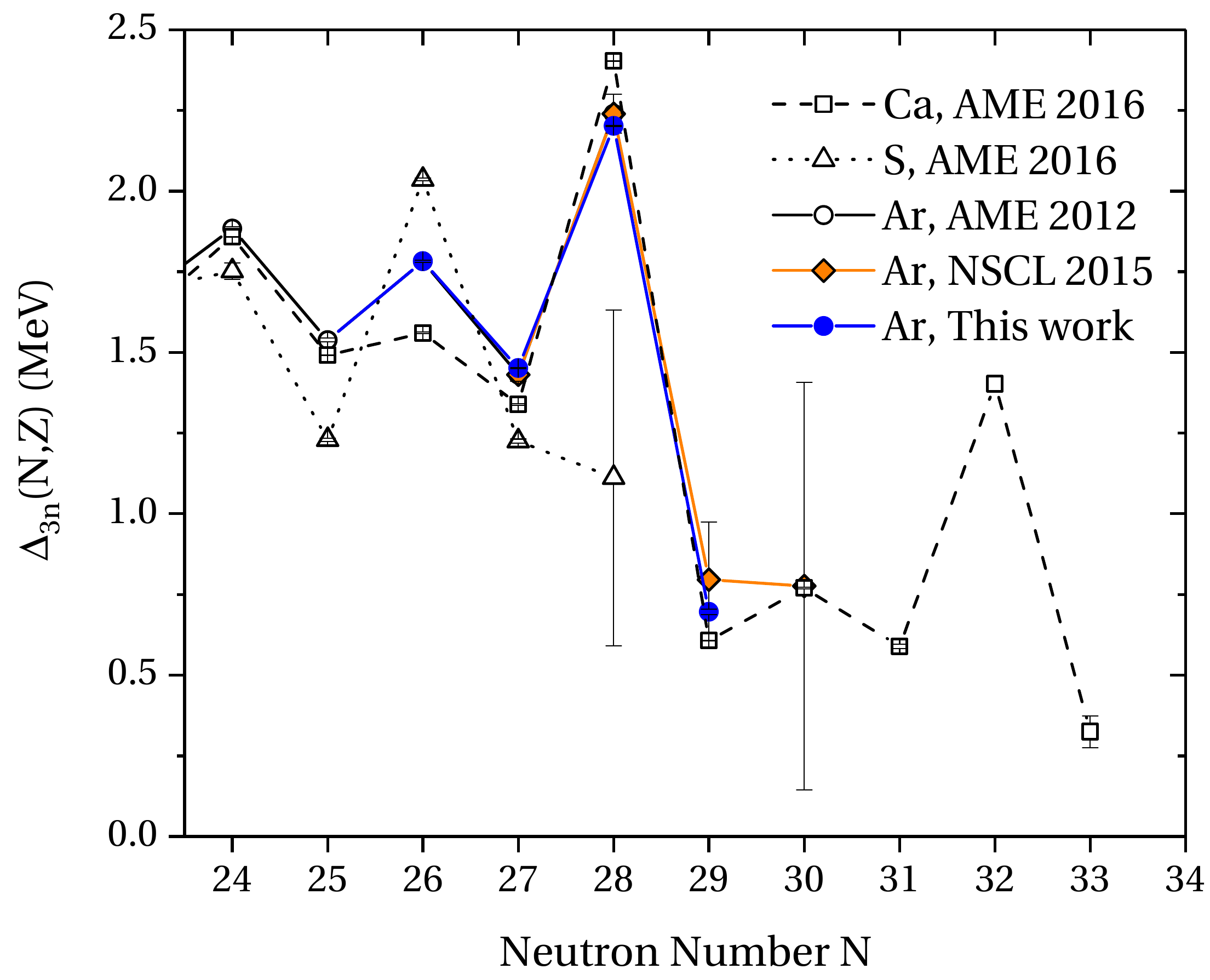}
\caption{Three-point estimator of the pairing gap for the calcium, argon and sulfur ($Z$~=~20, 18, 16, respectively) isotopic chains. The calcium and sulfur values are extracted from the AME2016 \cite{AME2016} and are represented as open square and triangles, respectively. For the argon isotopic chain, values extracted from AME2012 \cite{AME2012} are represented as open circles while the orange diamonds represent the trend obtained from the NCSL 2015 measurements \cite{Meisel2015}. The trend obtained from this work is represented as blue circles.}
\label{d3n_exp}
\end{figure}

In order to examine the strength of the empirical shell gap at \emph{N} = 28 more directly, Fig.~\ref{d3n_exp} shows another mass filter, namely the three-point estimator of the pairing gap, defined as $\Delta_{3n}(N,Z) =  \frac{(-1)^{N}}{2} \left[ ME(Z,N+1) - 2 ME(Z,N) + ME(Z,N-1) \right]$. This quantity is usually discussed in the context of the study of the odd-even staggering of binding energies, but at the crossing of a neutron-shell closure $N_{0}$ this staggering is enhanced and 
$\Delta_{3n}(N_{0},Z)$ is then directly related to the one-neutron empirical shell gap following: $\Delta_{1n}(N_{0},Z) = S_{1n}(N_{0},Z)-S_{1n}(N_{0}+1,Z) = 2\times \Delta_{3n}(N_{0},Z)$. The strength of the empirical one-neutron shell gap in $^{46}$Ar estimated from this work is $\Delta_{1n}(28,18) =$ 4.405(4)~MeV. This value is in agreement with that obtained from the study of the $^{46}$Ar(d,p)$^{47}$Ar reaction \cite{Gaudefroy2006}. As a result, even if all the masses measured in this work are found to be more bound than in \cite{AME2012,Meisel2015}, they reveal a net reduction of the $N$ = 28 one-neutron empirical shell-gap in the argon chain by 73~keV. In addition, compared to $^{48}$Ca, $^{46}$Ar exhibits a $N$ = 28 shell gap which is 402(4)~keV smaller (see Fig.~\ref{d3n_exp}). Given the doubly magic character of $^{48}$Ca, investigating only the systematics of the mass surface, one would conclude that the $N$ = 28 shell is a quite robust shell closure down to $Z$ = 18, thus confirming the findings of \cite{Meisel2015}. On the contrary, the demise of the $N$ = 28 shell closure in the sulfur chain is suggested by the strong reduction of the one-neutron shell gap between $Z$ = 18 and $Z$ = 16, although the large uncertainty calls for precision mass measurements.

\begin{figure}
\centering
\includegraphics[scale=0.35]{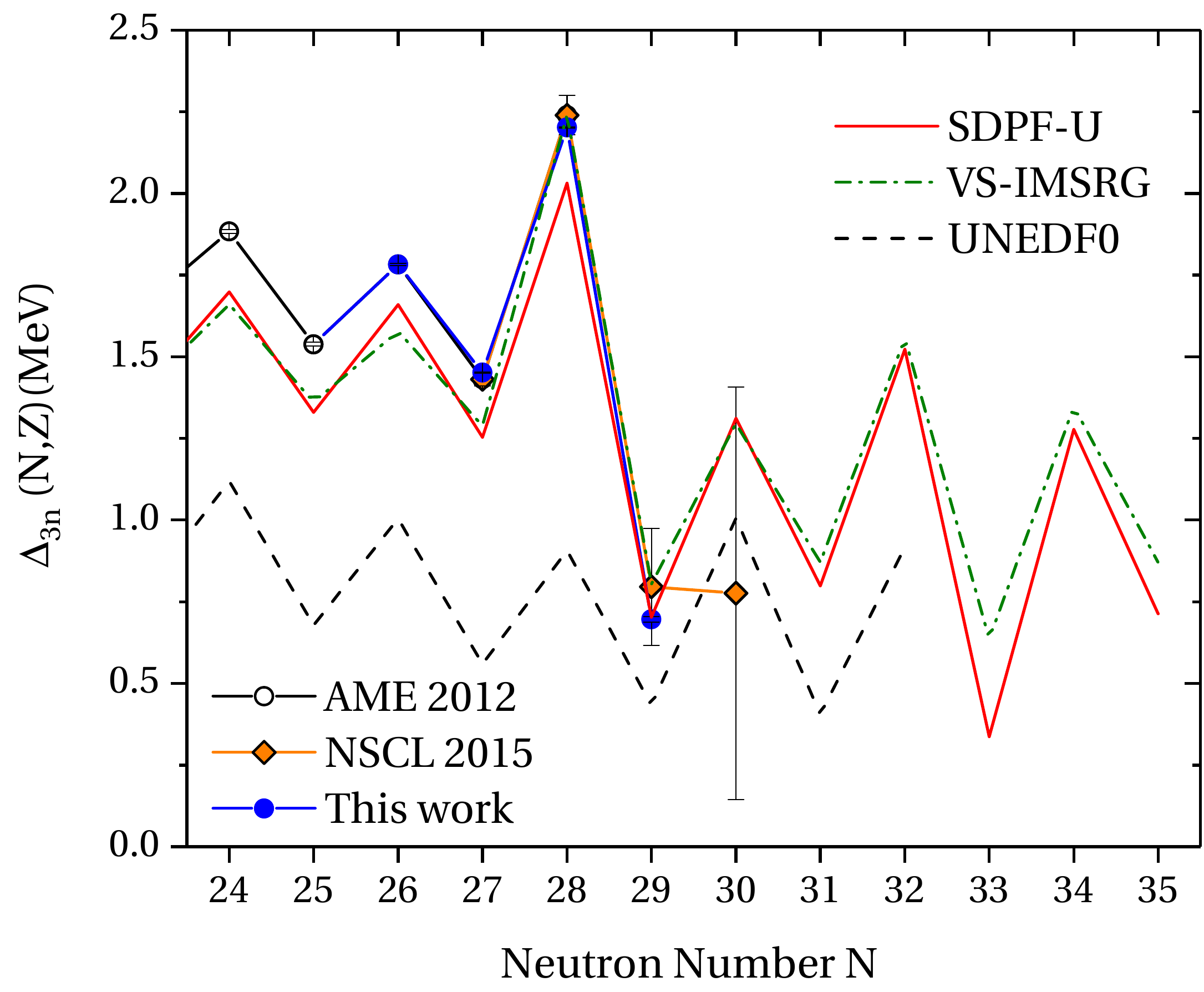}
\caption{Comparison between the three-point estimator of the pairing gap for the argon chain obtained from this work and the ones predicted from the UNEDF0 density functional, $SDPF-U$ shell model and the {\it ab initio} VS-IMSRG.}
\label{d3n_theo_ar}
\end{figure}

In order to gain further insights into the physics at play within this region of the nuclear chart, the binding energy trends obtained in this work were confronted with predictions from various theoretical approaches. To this end, mean-field calculations of even-even and odd-even argon isotopes were performed using the UNEDF0 energy-density functional \cite{PhysRevC.82.024313}. For these calculations a surface-volume-type pairing interaction was chosen. Its strength was kept fixed, since the UNEDF0 functional simultaneously fits this with the other functional parameters. The HFBTHO code, which solves the HFB equations enforcing axial symmetry \cite{STOITSOV20131592}, was used. The odd-\emph{N} isotopes were computed performing quasi-particle blocking within the so-called equal-filling approximation \cite{PhysRevC.78.014304}. The Lipkin-Nogami prescription was used for approximate particle-number restoration. The obtained trend of $\Delta_{3n}(N,Z)$ is presented in Fig.~\ref{d3n_theo_ar}. A first observation is that none of the characteristic features indicative of shell-closure at $N$ = 28 are reproduced. Furthermore, the overall scale of the predicted $\Delta_{3n}(N,Z)$ trends is greatly underestimated. This indicates that the adjusted UNEDF0 pairing strength is too weak to correctly describe this region of lighter nuclides. 

The spectroscopic results in this region are believed to be well understood within the framework of the phenomenological shell model \cite{Gaudefroy2006,PhysRevC.81.064329,Bhattacharyya2008}. Thus, calculations were performed using the \emph{m-scheme} shell-model code ANTOINE \cite{antoine1,antoine2} using the \emph{SDPF-U} shell-model interaction \cite{PhysRevC.79.014310}. In the calculation, the neutron valence space spans the entire \emph{sd-pf} shell, while protons are restricted to the \emph{sd} shell. An additional constraint is that particle excitations between the \emph{sd} and \emph{pf} shells are forbidden (i.e., a so-called $0\hbar \omega$ calculation). 

The trend of $\Delta_{3n}(N,Z)$ obtained from the calculated argon ground states is shown in Fig.~\ref{d3n_theo_ar}. A 250-keV offset notwithstanding, the agreement between theory and experiment is excellent, highlighting the ability of the \emph{SDPF-U} interaction to not only reproduce spectroscopy along the argon isotopic chain \cite{PhysRevC.81.064329}, but also  binding-energy systematics.

\begin{figure}
\centering
\includegraphics[scale=0.35]{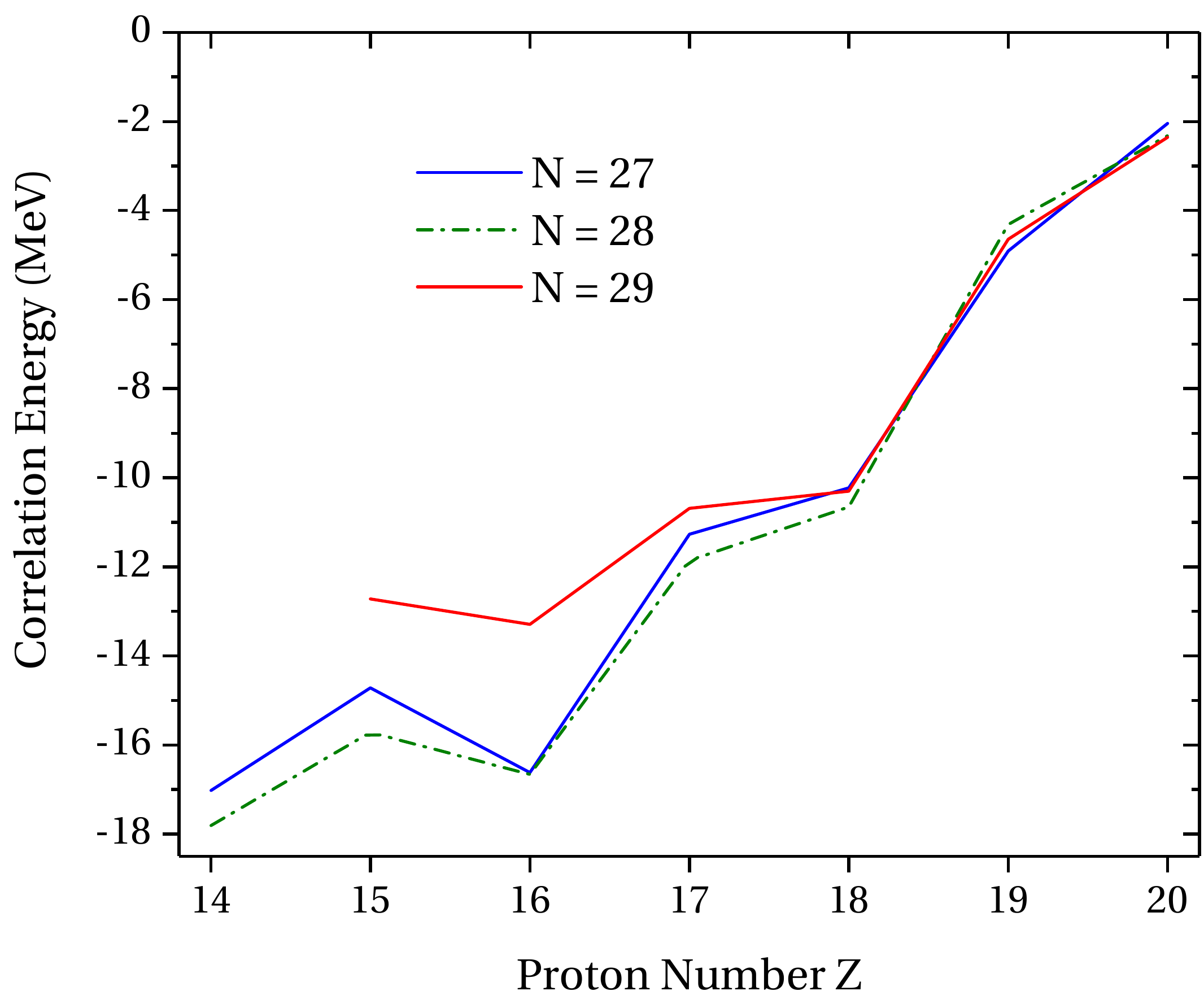}
\caption{Evolution of the ground-state correlation energy calculated using the \emph{SDPF-U} shell-model interaction \cite{PhysRevC.79.014310} as a function of the proton number $Z$ for the \textit{N} = 27 (dashed blue line), 28 (dash-dotted green line), 29 (solid red line) isotones.}
\label{cor_energy}
\end{figure}

The presence of a strong shell-closure at $N$ = 28 should be characterized by the predominance of the $\nu (1f7/2)^{8}$ \emph{natural} configuration in the ground-state wavefunction of even Ar isotopes. A so-called \emph{intruder} configuration would be characterised by the promotion of at least one such $1f7/2$ neutron to higher energy orbitals. 
Hence, in agreement with \cite{Bhattacharyya2008}, our calculations show that the ground-state of the doubly-magic $^{48}$Ca isotope is built at $\approx$90 \% on the \emph{natural} configuration while the ground-states of $^{46,48}$Ar is only built at $\approx$50 \% on this same configuration. In addition, the \emph{monopole} and \emph{multipole} energy contributions of the calculated ground-state energies were extracted. While the \emph{monopole} energy represents single-particle contributions, of spherical Hartree-Fock type, the \emph{multipole} energy was shown to represent the contribution of correlations to the total energy of a calculated shell-model state \cite{Dufour1996}. The evolution of the calculated ground-state correlation energy for $Z$ = 14-20 and $N$ = 27-29 is shown in Fig. \ref{cor_energy}. Hence, in agreement with \cite{PhysRevC.81.064329}, we find a rapid increase of correlation energy south of $^{48}$Ca. In $^{48,49}$Ca, correlations account for $\approx$2~MeV of the total energy of the ground state. On the contrary, for $^{46,47}$Ar the correlation energy is already $\approx$11~MeV, when only two protons are removed from the closed calcium proton core. In comparison, the measured strength of the one-neutron empirical shell gap is close to $\approx$4.8~MeV and  $\approx$4.4~MeV for $^{48}$Ca and $^{46}$Ar respectively. As a result, in agreement with previous shell-model studies performed with the phenomenological \emph{SDPF-U} interaction, we find that the ground-states of the studied argon isotopes do not exhibit the expected characteristics of a typical closed-shell nucleus, but rather suggests that collectivity is already emerging only two protons below $^{48}$Ca. 

This observation establishes the argon chain as the transitional point from the closed-shell region around calcium towards a region of collectivity below $Z$ = 18. This conclusion is also supported by other experimental evidence \cite{Gade2005,Bhattacharyya2008}, the most compelling of which is the spectroscopic factor from a $^{46}$Ar(d, p)$^{47}$Ar transfer reaction \cite{Gaudefroy2006}. Indeed, this reaction populates a $7/2^{-}$ state in $^{47}$Ar for which the model-dependent determined vacancy is 1.36(16). Again, this is in contradiction with the expectations of a naive shell-model picture of a closed-shell $^{46}$Ar. As a result, the conclusion drawn from the mass systematics alone of a strong shell closure in $^{46}$Ar \cite{Meisel2015} must be nuanced in light of the wealth of experimental and theoretical data.

\begin{figure}
\centering
\includegraphics[scale=0.35]{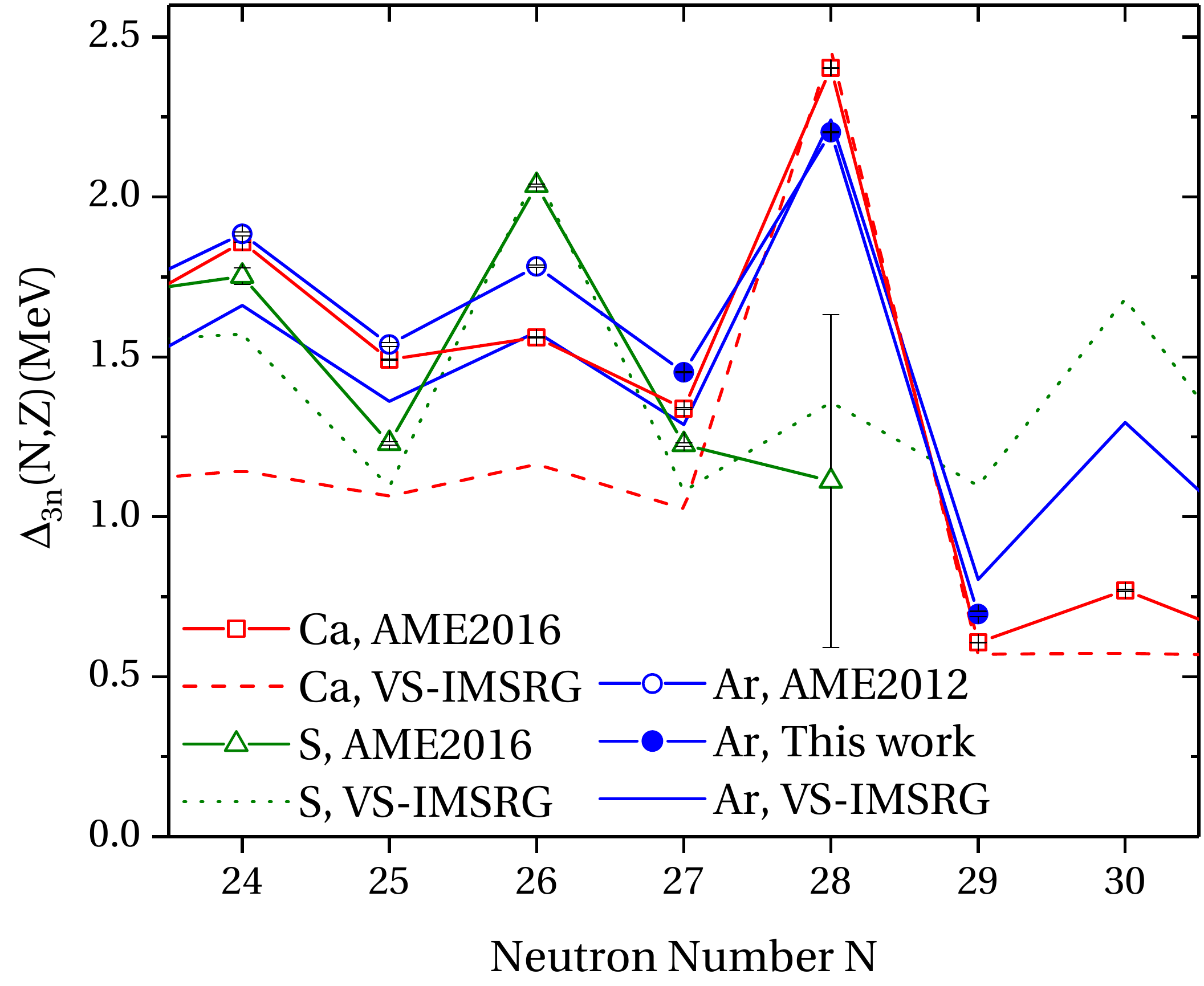}
\caption{Comparison between the empirically determined pairing-gap trend and the one obtained in VS-IMSRG calculations for the calcium (red), argon (blue) and sulfur chains (green). For the calcium and sulfur isotopic chains the experimental values from AME2016 \cite{AME2016} are represented as open squares and triangles, respectively. For the argon isotopic chain, values extracted from AME2012 \cite{AME2012} are represented as open circles while the plain blue circles are values from this work. The VS-IMSRG predictions are represented as dashed, dotted and solid lines for the calcium, sulfur and argon chains, respectively.}
\label{d3n_reg}
\end{figure}

The ground states of the measured argon isotopes were also examined using the {\it ab initio} VS-IMSRG approach \cite{Tsuk12SM,Bogn14SM,Stro16TNO,Stro17ENO,Stro19ARNPS}. The spectroscopic quality of this approach has been recently studied in light of the first measurement of the $2_{1}^{+}$ state in $^{52}$Ar \cite{Liu2019}. While the \emph{SDPF-U} phenomenological interaction provided the best overall description of the evolution of the $2_{1}^{+}$ states along the argon chain, the VS-IMSRG approach nonetheless reasonably well reproduced this trend up to $^{52}$Ar. In this work we start from the 1.8/2.0 (EM) NN+3N interactions developed in \cite{Hebe11fits,Simo17SatFinNuc}, which reproduces the ground-state energy systematics, including the location of the proton and neutron driplines, of nuclei throughout the light to medium-mass regions \cite{Simo16unc,Hag16,Rui16,Simo17SatFinNuc,Holt2019}. Details of the calculations are the same as those given in \cite{Simo17SatFinNuc}, unless explicitly stated otherwise. In particular, we use the Magnus formulation of the IMSRG \cite{Morr15Magnus,Herg16PR} to construct an approximate unitary transformation to first decouple the $^{28}$O core energy, then a proton $sd$ and neutron $pf$ valence-space Hamiltonian from the full $A$-body problem. In addition, with the ensemble normal-ordering procedure of Ref.~\cite{Stro17ENO}, we approximately include effects of 3N forces between valence nucleons, such that a specific valence-space Hamiltonian is constructed for each nucleus to be studied. The final diagonalization is performed using the NuShellX@MSU shell-model code~\cite{BROWN2014115}. 

\begin{figure}
\centering
\includegraphics[scale=0.35]{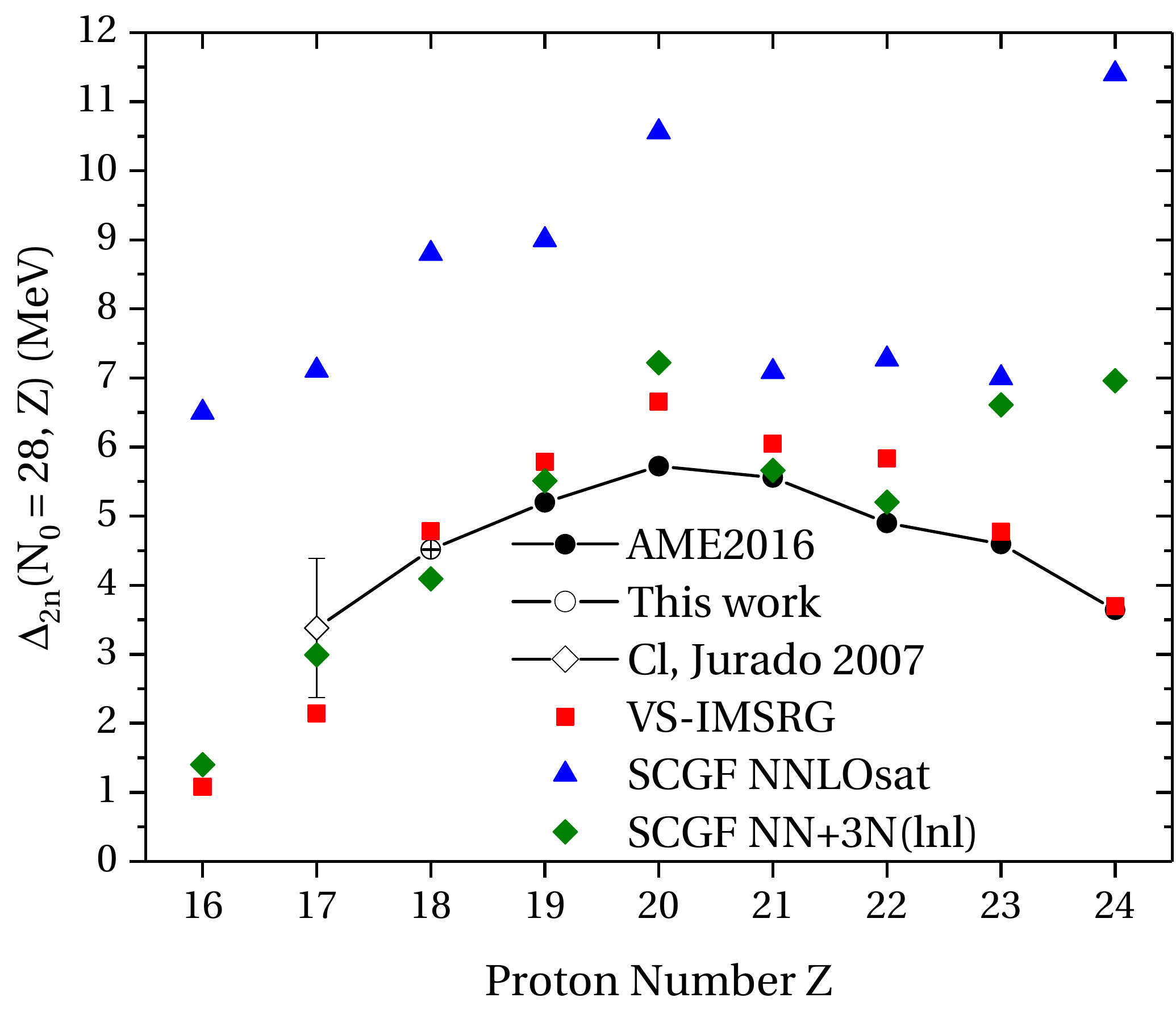}
\caption{$N$ = 28 two-neutron empirical shell gap for elements ranging from sulfur to chromium. Experimental values are represented as black circles \cite{AME2016}. The open circle represents the value from this work while the open diamond represents the value extracted from \cite{Jurado2007} (not included in the AME evaluation).}
\label{d2n}
\end{figure}

The trend of $\Delta_{3n}(N,Z)$ obtained from these calculations is also shown in Fig.~\ref{d3n_theo_ar}, revealing that the experimental $\Delta_{3n}(N,Z)$ trend is also very well reproduced along the entire argon chain, particularly the magnitude of the one-neutron empirical shell gap. Figure~\ref{d2n} shows the $N$ = 28 two-neutron shell gap, defined as $\Delta_{2n} = S_{2n}(N,Z) -S_{2n}(N+2,Z)$, obtained from various theoretical approaches as a function of $Z$. The VS-IMSRG prediction for the two-neutron shell gap at $Z$~=~18 is also in good agreement with the one obtained from the masses measured in this work, despite modestly overestimating it by $\approx$500~keV. The $N$~=~28 two-neutron gap for the calcium chain is however overestimated by more than 1~MeV. Nonetheless, we see that the \emph{ab initio} approach of the VS-IMSRG offers a consistent framework for predicting the systematics of ground- and excited-state energies simultaneously throughout the argon chain.

We also examined the composition of the wavefunctions obtained within the VS-IMSRG approach. In complete analogy with the conclusions drawn from the phenomenological \emph{SDPF-U}-interaction, we find that the ground state of $^{46}$Ar is not majoritarily ($\approx$40 \%) built on the \emph{natural} $\nu (1f7/2)^{8}$ configuration while the ground state of the benchmark doubly closed-shell $^{48}$Ca nucleus is built at $\approx$90 \% on that same configuration. In addition, to assess the quality of the VS-IMSRG prediction in this transitional region, we also perform calculations in the sulfur isotopic chain. The trend of $\Delta_{3n}(N,Z)$ obtained from these calculations is shown in Fig.~\ref{d3n_reg}. Here we see that not only the magnitude of the empirical one-neutron shell gaps in both $^{48}$Ca and $^{46}$Ar are well reproduced, but also that the erosion of the $N$ = 28 shell closure, as extracted from the mass systematics in the sulfur chain \cite{PhysRevLett.84.5062,Jurado2007,Ringle2009}, emerges \emph{ab initio}. The marked reduction of the predicted $N$ = 28 two-neutron shell gap from $Z$~=~18 to $Z$~=~16 is apparent also in Fig.~\ref{d2n}. Therefore a precise determination of the $^{45,46}$S masses is highly desirable in order to firmly assess the agreement between theory and experiment. While a systematic study of the entire region is beyond the scope of the present article, the VS-IMSRG offers a promising and consistent framework to guide future experimental efforts in the region of deformation below $^{48}$Ca.

To complete our \emph{ab initio} analysis, many-body calculations within the Gorkov-SCGF approach \cite{PhysRevC.84.064317,Soma14a} were performed for closed- and open-shell isotopes around $N$~=~28 and with $Z = 16-24$.
Medium-mass nuclei around $Z$ = 20 had been previously investigated within this framework \cite{PhysRevC.89.061301,Rosenbusch2015} using the NN+3N(400) chiral Hamiltonian of Refs.~\cite{PhysRevLett.109.052501,PhysRevC.68.041001,Navratil2007}. 
That study had revealed a satisfying reproduction of the binding-energy trend (namely two-neutron separation energies) for the Ca, K and Ar chains, although the agreement with experiment was worsening when going south of the Ca chain. 
The calculations are extended here using two more recent Hamiltonians. 
The first such interaction is the NNLO$_{\text{sat}}$~\cite{Ekstrom2015}, which 
departs from the traditional strategy of fitting to only few-body systems, and also includes observables up to $A$ = 25.
This procedure allows to correct for the poor saturation properties of the original NN+3N(400) Hamiltonian and leads to a reasonable reproduction of binding energies and charge radii up to the nickel chain~\cite{som2019}.
Another Hamiltonian labelled NN+3N(lnl) has been proposed to remedy some of the fundamental shortcomings of the NN+3N(400).
Contrarily to NNLO$_{\text{sat}}$, NN+3N(lnl) is adjusted solely on systems with $A$ = 2, 3 and 4.
First benchmark calculations on O, Ca and Ni chains~\cite{som2019} as well as application to K and Ca isotopes~\cite{PhysRevLett.120.062503, Chen19, Sun20} indicate that it constitutes a valuable addition to existing chiral Hamiltonians. 

The results obtained for elements with $Z = 16-19$ and $Z = 21-24$ with these new Hamiltonians are presented here for the first time. Calculations were performed in a spherical harmonic-oscillator basis including up to 14 major shells (e$_{max}$ = 13) while the three-body matrix elements were restricted to e$_{3max}$ = 16 $<$ 3e$_{max}$. A fixed oscillator frequency $\hbar \Omega $ = 20 MeV was used for the NNLO$_{\text{sat}}$ Hamiltonian, while $\hbar \Omega $ = 18 MeV was chosen for NN+3N(lnl). These correspond to the optimal values for total binding energies in this mass region~\cite{som2019}.

SCGF results with these two Hamiltonians for the $N$~=~28 two-neutron shell-gap as a function of the proton number $Z$ are displayed in Fig.~\ref{d2n}.
First, we observe that both interactions predict the emergence of the $N$~=~28 shell closure in $^{48}$Ca and its progressive demise in $^{46}$Ar and $^{44}$S.
Nonetheless, a marked difference between the SCGF-NN+3N(lnl) and SCGF-NNLO$_{\text{sat}}$ values is seen. The latter generally overestimates the strength of the two-neutron gap by several MeV, while the former offers a level of agreement with experimental data comparable to that of the VS-IMRSRG. 
It is noteworthy that both the VS-IMSRG and SCGF-NN+3N(lnl) approaches predict a two-neutron gap in $^{44}$S of similar magnitude. 
Above $Z$ = 20, SCGF calculations first follow the experimental trend displaying a decrease of the gap for scandium and titanium, then depart from experimental data for vanadium and chromium.
This disagreement signals the deterioration of the accuracy for doubly open-shell systems that display significant deformation.
Indeed, at present the Gorkov-SCGF framework achieves an efficient treatment of pairing correlations by breaking the U(1) symmetry associated to particle number, but enforces conservation of rotational symmetry, which leads to an inefficient account of deformation.
While this approach allows to tackle a large number of open-shell systems that do not exceedingly depart from sphericity, it looses accuracy when quadrupole correlations play a major role, which is presumably the case for nuclei like $^{49}$V and $^{50}$Cr.
The fact that this effect is not seen for sulfur and chlorine isotopes does not contradict the findings of the shell-model calculations, but rather points to a more mild impact of collectivity in those nuclei, at least for the description of the ground states.

\section{Conclusion}

In summary, we performed high-precision measurements of the atomic masses of $^{46-48}$Ar using the ISOLTRAP mass spectrometer at ISOLDE/CERN. Despite severe stable molecular contamination, the masses of $^{46-47}$Ar were successfully measured using the ToF-ICR method in a Penning trap, while the mass of $^{48}$Ar was determined by use of MR-ToF mass spectrometry. No statistically significant deviations were found when compared to literature values, but the uncertainties were reduced by up to a factor 90. The trends of nuclear binding energies obtained from the measured masses were used to probe the $N$ = 28 shell closure in neutron-rich argon isotopes. The systematics of the one- and two-neutron shell gaps indicate the presence of a persistent, yet reduced empirical shell gap in $^{46}$Ar compared to the doubly magic $^{48}$Ca, in accordance with results of previous mass measurements. More specifically, the one-neutron empirical shell gap is found to be reduced by only 402(4)~keV between $^{48}$Ca and $^{46}$Ar. However, taking into account the wealth of spectroscopic data available and using shell-model calculations performed with the \emph{SDPF-U} interaction, this conclusion must be nuanced. Indeed, $^{46}$Ar is found to form a transition point between the doubly closed-shell $^{48}$Ca and the collective $^{44}$S ground state.

A theoretical investigation of the measured isotopes was also performed using state-of-the-art \emph{ab-initio} approaches. The VS-IMSRG calculations reproduce the ground-state energy behavior in the argon chain as well as the phenomenological \emph{SDPF-U} interaction, thus providing an \emph{ab initio} description of the underlying physics in this region. 
SCGF calculations were also performed using two different Hamiltonians, NNLO$_{\text{sat}}$ and the recently derived NN+3N(lnl). 
Also in this case a progressive reduction of the  empirical two-neutron shell-gap was observed from $Z$ = 20 to $Z$ = 16.
While SCGF-NNLO$_{\text{sat}}$ results overestimate the strength of the two-neutron shell gap at ($Z,N$)=(18, 28), SCGF-NN+3N(lnl) closely follow those obtained from the VS-IMSRG, confirming the very good performance of the NN+3N(lnl) interaction in this mass region.

Accurate mass measurements extending the present study to more neutron-rich argon isotopes approaching $N$ = 32, 34 and to the sulfur isotopes beyond $N$ = 28 are highly desirable to put the predictions from the presented \emph{ab initio} approaches to the test. To this end, the present mass values constitute ideal anchor points for future experimental campaigns reaching further away from stability. 

\begin{acknowledgments}
M.M and D.L thank L. Gaudefroy for fruitful discussions which helped improve this article. We thank the ISOLDE technical group and the ISOLDE Collaboration for their assistance. We acknowledge support from the Max Planck Society, the German Federal Ministry of Education and Research (BMBF) (Contracts No.~05P12HGCI1, 05P15ODCIA, 05P15HGCIA and 05P18RDFN1), the Deutsche Forschungsgemeinschaft (DFG, German Research Foundation) -- Project-ID 279384907 -- SFB 1245, the French IN2P3, the United Kingdom Science and Technology Facilities Council (STFC) (Grants No. ST/P005314/1 and No. ST/L005816/1) and the European Union’s Horizon 2020 research and innovation programme (Grant No. 654002). J.K. acknowledges support from the Wolfgang Gentner Ph.D. scholarship (Grant No. 05E12CHA). Computations were performed at the J\"ulich Supercomputing Center (JURECA). SCGF calculations were performed by using HPC resources from GENCI-TGCC, France (Contract No. A007057392) and the DiRAC DiAL system at the University of Leicester, UK (BIS National E-infrastructure Capital Grant No. ST/K000373/1 and STFC Grant No. ST/K0003259/1).
\end{acknowledgments}

\bibliography{thesis_ar_cr}

\end{document}